\documentclass[10pt,prd, %preprint,
aps,
amsmath,
amssymb,
twocolumn,
nofootinbib,
superscriptaddress,
]{revtex4-1}
\usepackage{bm}
\SetSymbolFont{largesymbols}{bold}{OMX}{txex}{b}{n}
\usepackage{calc}
\usepackage{slashed}
\usepackage{tabularx}
\usepackage[export]{adjustbox}
\usepackage[english]{babel}
\usepackage{hyperref}
\usepackage{url}
\renewcommand{\vec}[1]{\bm{#1}}
\begin{document}
\title{Resummation of QED radiative corrections in a strong constant crossed field}
\author{A.A.~Mironov}\email{mironov.hep@gmail.com}
\affiliation{Prokhorov General Physics Institute of the Russian Academy of Sciences, Vavilova str. 38, Moscow, 119991, Russia}
\affiliation{National Research Nuclear University MEPhI, Kashirskoe sh. 31, Moscow, 115409,  Russia}
\affiliation{Steklov Mathematical Institute, Russian Academy of Sciences, Gubkina str. 8, Moscow, 119991, Russia}
\author{S.~Meuren}\email{smeuren@stanford.edu}
\affiliation{Department of Astrophysical Sciences, Princeton University, Princeton, NJ 08544, USA}
\affiliation{Stanford PULSE Institute, SLAC National Accelerator Laboratory, Menlo Park, CA 94025, USA}
\author{A.M.~Fedotov}\email{am\_fedotov@mail.ru}
\affiliation{National Research Nuclear University MEPhI, Kashirskoe sh. 31, Moscow, 115409,  Russia}
\affiliation{Laboratory for Quantum Theory of Intense Fields, National Research Tomsk State University, 
	Lenin Ave. 36, Tomsk, 634050, Russia}

\begin{abstract}
By considering radiative corrections of up to 3rd-loop order, Ritus and Narozhny conjectured that the proper expansion parameter for QED in a strong constant crossed field is $g=\alpha\chi^{2/3}$, where the dynamical quantum parameter $\chi=e\sqrt{-(Fp)^2}/m^3$ combines the particle momentum $p$ with the external field strength tensor $F$. Here we present and discuss the first non-perturbative result in this context, the resummed bubble-type polarization corrections to the electron self-energy in a constant crossed field. Our analysis confirms the relevance of the scaling parameter $g$ to the enhancement of bubble-type radiative corrections. This parameter actually represents the characteristic value of the ratio of the 1-loop polarization bubble to the photon virtuality. After an all-order resummation we identify and discuss two contributions to the self-energy with different formation regions and asymptotic behavior for $g\gg1$.  Whereas the breakdown of perturbation theory occurs already for $g\gtrsim1$,  the leading-order result remains dominant until the asymptotic regime $g\gg 1$ is reached. However, the latter is specific to processes like elastic scattering or photon emission and does not have to remain true for general higher-order QED processes. 
% for non-perturbative effects induced by bubble-type radiative corrections.
%The developed tools are likely to be useful for obtaining further insights into the non-perturbative regime $g\gtrsim1$ in general and, in particular, for deriving definite experimental predictions.
\end{abstract}

%\pacs{
%	52.27.Ep, %Electron-positron plasmas
%	52.50.Dg, %Plasma sources
%	41.75.Jv,   %Laser-driven acceleration
%	42.50.Ct   %Quantum description of interaction of light and matter; related experiments
%}
%\keywords{high intensity laser facilities, focused laser pulses, QED cascades, $e^+e^-$ production}
\maketitle

\section{Introduction}
\label{sec:intro}

Strong electromagnetic fields show up in atomic physics \cite{ullmann2017high} (including heavy ion collisions \cite{rafelski2017probing} and passage of ulrarelativistic particles through crystals \cite{wistisen2018experimental}), astrophysics of compact objects \cite{cerutti2017electrodynamics}, at the interaction point of future lepton colliders \cite{yokoya1992beam}, and during the interaction of high-power lasers with matter \cite{mourou2011whitebook}. A strong field is often well described by a coherent state that is not significantly altered by the quantum processes which it facilitates. This justifies the strong field approximation, which originated in the works of Furry \cite{furry1951bound}, Sokolov and Ternov \cite{sokolov1952quantum}, and Keldysh \cite{keldysh1958effect}. Accordingly, one neglects quantum fluctuations and back-reactions on the field itself, and treats the field as an external, i.e., given, classical one. However, its impact on the quantum processes in question is taken into account exactly.

A very important case is a constant crossed field (CCF), for which both field invariants are zero ($\vec{E}\cdot\vec{H}=0$ and $E=H$). This 'instantaneous' approximation is robust in many situations involving ultra-relativistic particles \cite{ritus1985quantum}. Already in the very first considerations of the basic QED processes of photon emission and pair photoproduction it was observed that asymptotically, for $\chi\gg1$, the probabilities scale as $g=\alpha\chi^{2/3}$ in a CCF, where\footnote{We use units such that $\hbar=c=\varepsilon_0=1$, electron mass and charge are denoted by $m$ and $-e$ respectively ($e>0$), and the signature of the Minkowski metric is $(+,-,-,-)$.} $\alpha=e^2/4\pi$ is the fine structure constant \cite{nikishov1964quantum}. The so called dynamical quantum parameter $\chi=(e/m^3)\sqrt{-(F_{\mu\nu}p^\nu)^2}$ measures the rest-frame field strength in units of the Schwinger critical field $F_0=m^2/e$ \cite{nikishov1964quantum}. Later, the same scaling was also found for the one-loop polarization \cite{narozhny1969propagation} and mass \cite{ritus1970radiative} radiative corrections, related by the optical theorem to the probability rates for pair production and photon emission, respectively.

After the consideration of radiative corrections up to 3rd loop order, it was conjectured that $g$ might replace $\alpha$ as an effective expansion parameter for QED in a strong CCF \cite{ritus1972radiative,morozov1975elastic,morozov1977elastic,narozhny1979radiation,narozhny1980expansion,ritus1985quantum}. Nowadays, this supposition is known as the Ritus-Narozhny conjecture \cite{fedotov2017conjecture}. Radiative corrections, which have been calculated for a CCF, are shown in Table~\ref{tab:known}. Note that the 2nd and 3rd loop contributions containing vertex corrections are missing, as they have not been calculated yet. However, they were believed to be subleading \cite{narozhny1979radiation,narozhny1980expansion} (the results presented in \cite{morozov1981vertex} seem to contradict this assumption and should therefore be reconsidered). Even though the leading-order results [see diagrams (1a) and (1b)] already indicate the importance of $g$ for the overall scaling of radiative corrections, it is not clear from the outset that this parameter also determines the importance of higher-order contributions and thus the breakdown of perturbation theory.

\begin{table*} %[h!]
	\caption{\label{tab:known}Known asymptotic scaling for radiative corrections in a CCF to the polarization operator (left) and the mass operator (right). For each diagram the row specifies the $\chi\gg1$ asymptotic behavior together with the corresponding source. The dominant scaling in $\chi$ is highlighted in bold for each loop order.}
	\renewcommand*{\arraystretch}{1.6}
	\begin{ruledtabular}
		\begin{tabularx}{\textwidth}{lccr}
			\multicolumn{4}{c}{\bf 1 loop} \\ [0.5ex]
			$\,$ &
			\multicolumn{1}{c}{
				\begin{tabularx}{0.38\textwidth}{lcXr}
					(1a) & \includegraphics[valign=c,width=3cm]{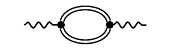}& $\alpha\bm{\chi^{2/3}}$ & \cite{narozhny1969propagation} \\ [2ex]
				\end{tabularx} 
			} &
			\multicolumn{1}{c}{
				\begin{tabularx}{0.38\textwidth}{lcXr}
					(1b) & \includegraphics[valign=c,width=3cm]{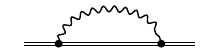}& $\alpha\bm{\chi^{2/3}}$ & \cite{ritus1970radiative} \\ [2ex]
				\end{tabularx}
			} &
			$\,$ \\ [1ex] \hline
			\multicolumn{4}{c}{\bf 2 loops} \\ [0.5ex]
			$\,$ &
			\multicolumn{1}{c}{
				\begin{tabularx}{0.38\textwidth}{lcXr}
					(2a) & \includegraphics[valign=c,width=3cm]{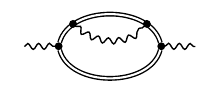}& \hspace*{-0pt}$\alpha^2\bm{\chi^{2/3}}\log{\chi}$ & \cite{morozov1977elastic} \\ [2ex]
				\end{tabularx}
			} &
			\multicolumn{1}{c}{
				\begin{tabularx}{0.38\textwidth}{lcXr}
					(2b) & \includegraphics[valign=c,width=3cm]{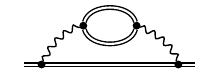} & $\alpha^2\bm{\chi}\log{\chi}$ & \cite{ritus1972radiative,ritus1972vacuum}\\
					(2c) & \includegraphics[valign=c,width=3cm]{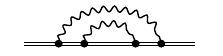} & $\alpha^2\chi^{2/3}\log{\chi}$ & \cite{morozov1975elastic} \\ [2ex]
				\end{tabularx}
			} &
			$\,$ \\ \hline
			\multicolumn{4}{c}{\bf 3 loops} \\ [0.5ex]
			$\,$ &
			\multicolumn{1}{c}{
				\begin{tabularx}{0.38\textwidth}{lcXr}
					(3a) & \includegraphics[valign=c,width=3cm]{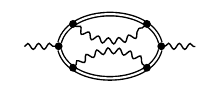}& $\alpha^3\chi^{2/3}\log{\chi}$ & \cite{narozhny1979radiation} \\
					(3b) & \includegraphics[valign=c,width=3cm]{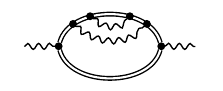}& $\alpha^3\chi^{2/3}\log{\chi}$ & \cite{narozhny1979radiation} \\
					(3c) & \includegraphics[valign=c,width=3cm]{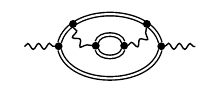}& $\alpha^3\bm{\chi}\log^2{\chi}$ & \cite{narozhny1980expansion} \\ [2ex]
				\end{tabularx}
			} &
			\multicolumn{1}{c}{
				\begin{tabularx}{0.38\textwidth}{lcXr}
					(3d) & \includegraphics[valign=c,width=3cm]{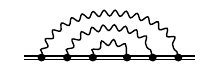}& $\alpha^3\chi^{2/3}\log^2{\chi}$ & \cite{narozhny1979radiation} \\
					(3e) & \includegraphics[valign=c,width=3cm]{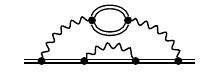}& $\alpha^3\chi^{4/3}$ & \cite{narozhny1979radiation} \\
					(3f) & \includegraphics[valign=c,width=3cm]{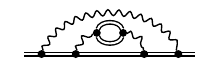}& $\alpha^3\chi\log^2{\chi}$ & \cite{narozhny1980expansion}\\
					(3g) & \includegraphics[valign=c,width=3cm]{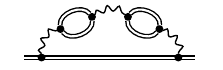}& $\alpha^3\bm{\chi^{5/3}}$ & \cite{narozhny1980expansion} \\ [2ex]
				\end{tabularx}
			} &
			$\,$ 
		\end{tabularx}	
	\end{ruledtabular}
\end{table*}

\begin{figure*}[tbp]
	\centering\includegraphics[width=.95\textwidth]{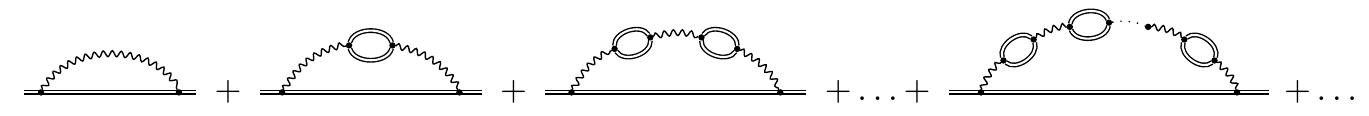}
	\caption{\label{fig:diagram} Bubble-type polarization corrections to the electron mass operator (double lines denote the dressed electron propagators in a constant crossed field \cite{schwinger1951gauge}). The corresponding exact photon propagator, obtained after resummation of the Dyson series with account for the 1-loop polarization operator, is referred to as the bubble-chain dressed photon propagator throughout the paper.}
\end{figure*}

To determine the effective expansion parameter, which governs the breakdown of perturbation theory, one has to examine the ratio between the $(n+1)$th and the $n$th loop order. For the mass operator (right column in Table~\ref{tab:known}) and $n=2$ this ratio is $\text{(3g)}/\text{(2b)}\sim g=\alpha \chi^{2/3}$. Narozhny conjectured that the same scaling will hold at all higher loop orders $n>2$ \cite{narozhny1980expansion}. The previously considered ratios $\text{(2b)}/\text{(1b)}\sim \alpha \chi^{1/3}\log\chi$ for the mass operator and $\text{(3c)}/\text{(2a)}\sim \alpha \chi^{1/3}\log\chi$ for the polarization operator initially caused some confusion about the correct expansion parameter \cite{ritus1972radiative,narozhny1979radiation}. The current interpretation is that these findings represent exceptions at the beginning of the expansion. Note that for the polarization operator these ratios are upshifted by one loop order with respect to those for the mass operator, as the polarization operator contains an extra fermion loop. It is therefore believed that $g$ also represents the effective expansion parameter of the polarization operator starting from $4$th-loop order, yet to be accurately calculated. The Ritus-Narozhny conjecture, as formulated in a final form in the paper\footnote{In fact, the assertions forming the conjecture are scattered along the concluding part of the paper \cite{narozhny1980expansion}, here we combine them all together.} \cite{narozhny1980expansion}, states that for $\chi\gg 1$: (i) the radiation probability and radiative corrections are enhanced by powers of $\chi$; (ii) the ratio of the dominant contributions to the $(n+1)$th and the $n$th orders of perturbation theory scales proportional to $g$ -- in this sense $g$ represents the effective expansion parameter for perturbation theory in a strong CCF; (iii) the corrections growing as the highest power of $g$ at each order of the perturbative expansion are those accommodating the maximal number of successive polarization loop insertions (bubbles) as shown in Fig.~\ref{fig:diagram}. 

Note that this is in sharp contrast to ordinary (field-free) QED, where the expansion parameter $\alpha$ is small and the effect of higher-order vacuum polarization corrections, after renormalization, is a logarithmic growth of the effective charge. As a result, polarization effects remain small for all reasonable energies, i.e., below the electro-weak unification scale. 

A situation which, at first glance, might appear very similar to a supercritical CCF, but which is actually qualitatively different, is the case of an electron/positron occupying the lowest Landau level (LLL) in a supercritical magnetic field \cite{loskutov1981behavior,gusynin1999electron}. In this case the applicability of dimensional reduction facilitates non-perturbative calculations, which have been carried out in the context of spontaneous chiral symmetry breaking (see e.g. \cite{gusynin1995dimensional}). The Ritus-Narozhny conjecture, however, applies to an ultra-relativistic electron/positron, which has quasi-classical trajectories \cite{tsai1974photon,tsai1975propagation,karbstein2013photon}. Thus, it effectively occupies very high Landau levels. Nevertheless, the LLL case can be mapped heuristically to the CCF case. To this end we note that for the ground Landau energy level $\varepsilon_{\text{LLL}}\propto \sqrt{B/F_0}$ \cite{berestetskii1982quantum}, the corresponding value $\chi\simeq (B/F_0)\times(\varepsilon_{\text{LLL}}/m)\simeq (B/F_0)^{3/2}$ effectively maps into $g\simeq\alpha B/F_0$ (c.f. \cite{shabad1975photon}). As to be expected, the two situations also exhibit qualitative differences. For example, the one-loop mass operator is only enhanced in a CCF \cite{ritus1970radiative,ritus1972radiative}, not in a supercritical magnetic field \cite{jancovici1969radiative}. 

Whereas supercritical magnetic fields are encountered in astrophysics, most researchers regarded a proof of the Ritus-Narozhny conjecture as an academic exercise with no practical relevance. This perspective has changed only recently, after realistic experimental proposals to probe the regime $g\gtrsim1$ were suggested. In particular, it was demonstrated that the value $g\simeq1$ can be attained by mitigation of rapid radiation losses in beam-beam collisions at a near-future lepton collider \cite{yakimenko2019prospect}. Alternatively, electrons could be collided with strong optical laser pulses at oblique incidence \cite{blackburn2019reaching} or head-on with strong attosecond pulses generated by reflection of high-power optical laser pulses from a solid target \cite{baumann2019probing}. Their passing through solid targets, which are irradiated from the back with ultraintense laser pulses, represents another suggested setup \cite{baumann2019laser}, as well as the channeling of multi-TeV electrons/positrons in aligned crystals \cite{di2019testing}.

\begin{figure}
	\begin{center}
		\includegraphics[valign=b,width=\columnwidth]{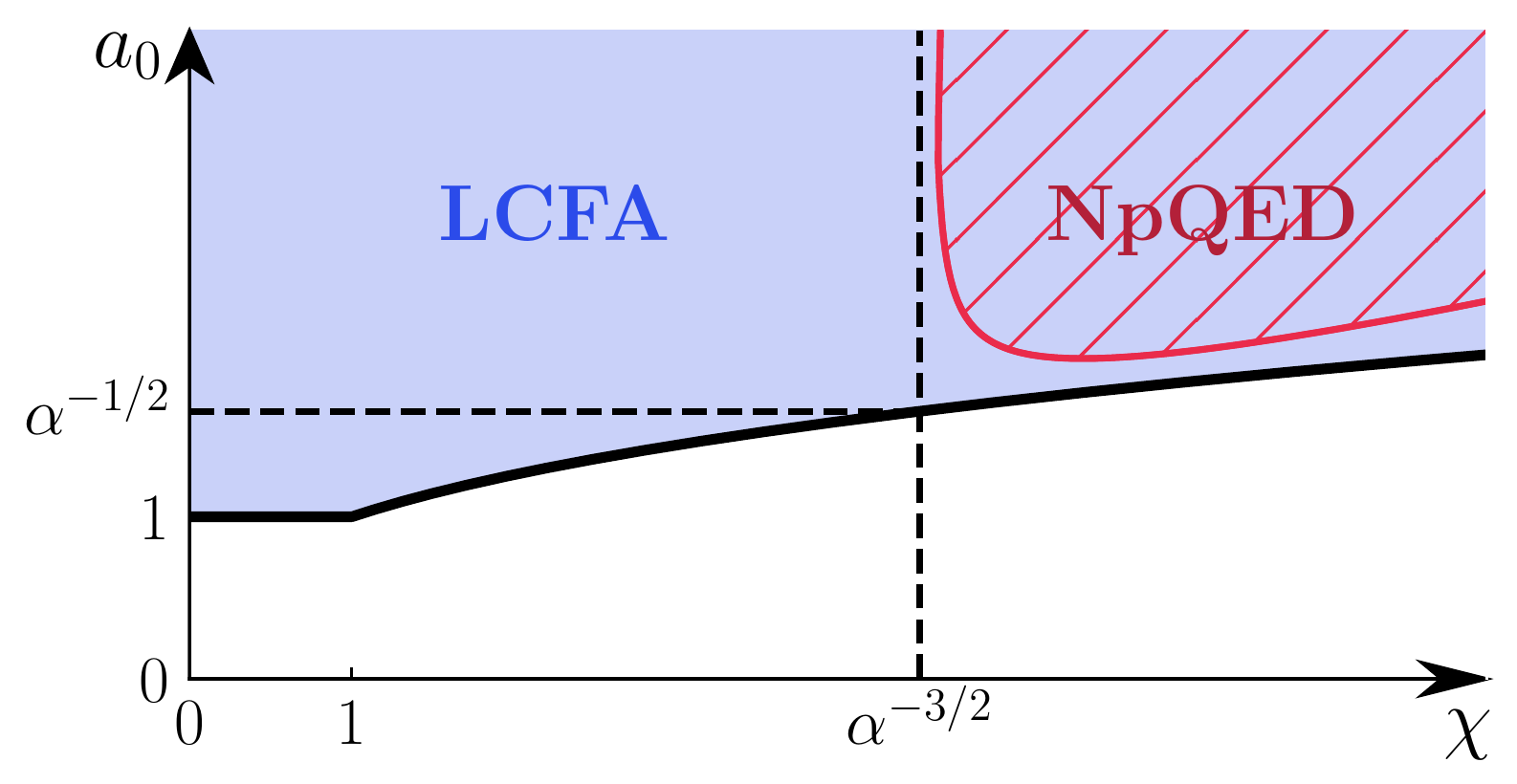}
	\end{center}
	\caption{Overview of the two most important parameters $(\chi,a_0)$ characterizing the interaction of a relativistic particle with a strong subcritical ($F\ll F_0$) field. The domain of validity of the locally constant field approximation (LCFA) $a_0\gg \max\{1,\chi^{1/3}\}$ is indicated in blue, and the subdomain of the non-perturbative regime $g=\alpha\chi^{2/3}\gtrsim1$ is hatched in red.}
	\label{fig:landscape}
\end{figure}

It is obvious that a CCF can be only approximately realized in practice. \label{lccf} According to recent discussions (see, e.g., \cite{harvey2015testing,blackburn2018benchmarking,di2018implementing} and the references therein) the locally constant field approximation (LCFA) is valid for describing scattering of ultra-relativistic particles in a strong subcritical ($F\ll F_0$) slowly varying field under the conditions $a_0\gg1$ and $a_0\gg\chi^{1/3}$, where $a_0=eF\tau/m$ is the classical non-linearity parameter. Here $F$ and $\tau$ are the typical field strength and field variation length/time, respectively. These conditions ensure that the typical formation scale for strong-field processes like photon emission, pair production or elastic scattering is smaller than the scale over which the field changes significantly. Under these conditions the results derived for a CCF are applicable \cite{yakimenko2019prospect}. Whereas the importance of the former condition ($a_0\gg1$) was realized and stated explicitly already in the initial publications on this topic (see, e.g., \cite{nikishov1964quantum,ritus1985quantum}), the necessity of the latter condition ($a_0\gg\chi^{1/3}$) was not widely known (previous works commonly implied $\chi\lesssim 1$), see \cite{baier1989quantum}, though. This is illustrated in Fig.~\ref{fig:landscape}, where the domain of validity of the LCFA is indicated in blue and the location of the non-perturbative regime is hatched in red. Recent rigorous considerations of the 1-loop mass and polarization operators in a strong pulsed field \cite{podszus2019high,ilderton2019note} explicitly demonstrated that in the high-energy limit, with field strength and duration kept fixed (given $a_0$), the scalings observed in a CCF no longer apply. Instead of a power law enhancement only a logarithmic scaling with $\chi$ is observed. This becomes obvious in Fig.~\ref{fig:landscape}. For fixed field strength and duration (given $a_0$) the high energy limit means a motion rightwards along a horizontal line. This inevitably implies that the domain of validity of the LCFA will be left. In fact, the effective charge exhibits a logarithmic dependence on the field strength parameter $\chi$ even in a pure CCF \cite{artimovich1990properties}. However, as we will show below, this is irrelevant to the Ritus-Narozhny conjecture, which focuses on the scaling of the effective masses.

Here we revisit the Ritus-Narozhny conjecture and present the first all-order resummation of the bubble-type polarization corrections to the electron self-energy shown in Fig.~\ref{fig:diagram}. According to the Ritus-Narozhny conjecture this should be the dominating contribution to the cumulative higher-order radiative corrections for $g\gtrsim 1$. Note that a similar resummation of the 1-loop radiative corrections to external electron and photon lines in a laser field was previously discussed in \cite{meuren2011quantum,meuren2015polarization}, see also \cite{meuren2015nonlinear} for more details. Our consideration not only confirms the importance of the parameter $g$ for such kind of corrections, but also provides further insights into its nature and importance. 

The rest of the paper is organized as follows. After introducing our notation and technical preliminaries in Section~\ref{sec:massop}, we discuss how the parameter $g$ emerges in bubble diagrams in Section~\ref{subsec:pert}. Next, in Section~\ref{subsec:appr}, we present an approximation which facilitates their all-order analytic resummation and identify two qualitatively different contributions, one associated with photon emission and another one related to trident pair production. Their explicit evaluation for $\chi\gg1$ is finalized in Section~\ref{sec:u_sigma}. A detailed summary and further discussion of our results and their implications are presented in Section~\ref{sec:summ}. To keep our presentation succinct, we summarize the main properties of the 1-loop polarization operator in a CCF in the Appendix.

\section{Bubble-type polarization corrections to the mass operator in a constant crossed field}
\label{sec:massop}

In this paper we focus on studying the bubble-type polarization corrections to the electron self-energy in a CCF (see Fig.~\ref{fig:diagram}), or, more precisely, to the on-shell elastic electron scattering amplitude $T_s(p)=-\mathcal{M}(\chi)/(2p^0)$, where the invariant amplitude $\mathcal{M}(\chi)\equiv \bar{u}_{p,\lambda} M u_{p,\lambda}$ depends on the dynamical parameter $\chi$. Here $M$ is the mass operator of an electron and $u_{p,\lambda}$ is a free Dirac spinor characterizing the electron spin state. 

In the Ritus $E_p$-representation \cite{ritus1970radiative,ritus1972radiative,ritus1985quantum} the correction to the mass operator in a CCF depicted in Fig.~\ref{fig:diagram} reads
\begin{align}\nonumber
-iM(p',p)=&\int d^4 x\, d^4 x'\,\bar{E}_{p'}(x')(ie\gamma^\mu)  \\ \nonumber
& \quad \times S^c(x',x)(ie\gamma^\nu) E_{p}(x) D^c_{\mu\nu}(x',x)\\ \nonumber
=&\int \frac{d^4 l}{(2\pi)^4} \frac{d^4 q}{(2\pi)^4} \Gamma^\mu(l;p',q) \\ 
&\quad \times \frac{i(\slashed{q}+m)}{q^2-m^2+i0} \Gamma^\nu(-l;q,p) D^c_{\mu\nu}(l). \label{gen_formula}
\end{align}
Here $S^c$ denotes the tree-level dressed electron propagator and $D^c$ is the bubble-chain dressed photon propagator \cite{narozhny1969propagation} attached to the electron line in Fig.~\ref{fig:diagram}. The 4-momenta of the virtual photon and electron in the outer loop are denoted by $l^\mu$ and $q^\mu$, respectively, $\slashed{q}=\gamma^\mu q_\mu$, and  $E_p(x)$ is a matrix solution to the Dirac equation in a CCF, which reduces to the unity matrix if the field is switched off adiabatically \cite{ritus1970radiative}. Furthermore,
\begin{equation}\label{dressed_vertex}
\Gamma^\mu(l;p,q)=\int d^4 x\,  e^{-ilx} \bar{E}_{p}(x)(ie\gamma^\mu) E_q(x)
\end{equation}
is called dressed vertex \cite{ritus1972radiative, mitter1975quantum}, where Dirac conjugation of a matrix $\bar{E}_{p}=\gamma^0E_p^\dagger\gamma^0$ is denoted by a bar. For the sake of clarity Eq.~(\ref{gen_formula}) is written in two different ways: the right-hand side of the upper line is written in a coordinate representation, whereas the lower line expresses the electron propagator in the $E_p$-representation and the photon propagator in the momentum representation.

The bubble-chain dressed photon propagator in a CCF reads \cite{narozhny1969propagation,ritus1972radiative,ritus1985quantum}
\begin{equation}\label{photon_prop}
D^c_{\mu\nu}(l)= D_0(l^2,\chi_l) g_{\mu\nu}+\sum\limits_{i=1}^2 D_i(l^2,\chi_l)\epsilon_\mu^{(i)}(l)\epsilon_\nu^{(i)}(l),
\end{equation}
where $\chi_l=(e/m^3)\sqrt{-(F_{\mu\nu}l^\nu)^2}$ is the dynamical quantum parameter of the virtual photon, $\epsilon_\mu^{(1)}(l)=eF_{\mu\nu}l^\nu/(m^3\chi_l)$ and $\epsilon_\mu^{(2)}(l)=eF^\star_{\mu\nu}l^\nu/(m^3\chi_l)$ are the normalized field-induced transverse 4-vectors, and $F^\star_{\mu\nu}=(1/2)\varepsilon_{\mu\nu\lambda\sigma}F^{\lambda\sigma}$ is the dual field strength tensor. The longitudinal component in Eq.~(\ref{photon_prop}) is given by
\begin{equation}\label{photon_prop_long}
D_0(l^2,\chi_l)=\frac{-i Z}{l^2+i0},
\end{equation}
and differs from the field-free one only by a finite factor $Z(l^2,\chi_l)$ [see Eq.~\eqref{Z}], whereas the transverse components
\begin{equation}\label{photon_prop_trans}
\begin{split}
D_{1,2}(l^2,\chi_l)&=\frac{i Z^2\Pi_{1,2}}{\left(l^2+i0\right)\left(l^2-Z\Pi_{1,2}\right)}\\
&=\frac{-i Z}{l^2+i0}-\frac{-i Z}{l^2-Z\Pi_{1,2}},
\end{split}
\end{equation}
exhibit additional poles corresponding to two effective photon masses (one for each transverse photon polarization state). They are determined by the renormalized eigenvalues $\Pi_{1,2}(l^2,\chi_l)$ of the polarization operator [see Eq.~\eqref{pi12}].

Overall, the only effect of the factor $Z(l^2,\chi_l)$ is to introduce an effective coupling $\alpha\mapsto \alpha_{\text{eff}}(l^2,\chi_l)=Z(l^2,\chi_l)\alpha$ (cf. \cite{artimovich1990properties}). However, $Z$ remains very close to unity for all reasonable values of $l^2$ and $\chi_l$. Therefore, we will ignore this logarithmic correction by setting $Z\approx1$ and $\alpha_{\text{eff}}\approx\alpha$ from now on. Further details are given in the Appendix. 

In the following we will simplify the expression obtained by combining Eqs.~\eqref{gen_formula}-\eqref{photon_prop_trans}. The part of the calculation which closely follows Ref.~\cite{ritus1972radiative} will only be outlined. Since $E_p(x)$ differs from a plane wave $e^{-ipx}$ only by a factor depending on $\varphi=kx$ ($k^\mu$ is directed along the Poynting 4-vector of the CCF, its normalization is arbitrary), the dressed vertex [see Eq.~\eqref{dressed_vertex}] in a CCF can be written in the following way
\begin{equation}\label{dressed_vertex_Fourier}
\Gamma^\mu(l;p,q)=\int_{-\infty}^{\infty} d\nu\,\delta^{(4)}(p-q-l-\nu k)\tilde{\Gamma}^\mu(\nu;p,q),
\end{equation}
where $\nu k^\mu$ is the energy-momentum transferred to the external field. $\tilde{\Gamma}^\mu(\nu;p,q)$ can be expressed in terms of the Airy function 
\cite{vallee2004airy}
\begin{equation}\label{Ai_def}
\mathrm{Ai}(t)=\frac{1}{2\pi}\int_{-\infty}^\infty d\sigma \, e^{-i(t \sigma+\sigma^3/3)}.
\end{equation}
Due to the transversality of a CCF the dresssed vertex remains invariant under translations of its arguments $p$ and $q$ by 4-vectors proportional to $k^\mu$.
The 4-dimensional $\delta$-function, shown explicitly in Eq.~(\ref{dressed_vertex_Fourier}), expresses energy-momentum conservation with the external CCF included \cite{ritus1972radiative}. Due to the presence of two such $\delta$-functions in Eq.~(\ref{gen_formula}) (one from each dressed vertex) $p'$ can actually differ from $p$ only by a 4-vector proportional to $k^\mu$. Hence we can apply the replacement $\tilde{\Gamma}^\mu(-\nu';p',q)\mapsto\tilde{\Gamma}^\mu(-\nu';p,q)$. Then, one of the two 4-dimensional $\delta$-functions in Eq.~\eqref{gen_formula} removes the integration over $d^4q$, after which only 6 integrations remain: over $d^4l$, $d\nu$ and $d\nu'$. 

It is convenient to apply the following changes of variables: $l^\mu \mapsto \lbrace l^2,\,u,\,\rho,\,\tilde\rho\rbrace$ and $\nu\mapsto \mu$, where $u=\chi_l/\chi_q$, $\chi_q$ is the dynamical quantum parameter of the electron in the outer loop, $\rho=p^\mu\epsilon_\mu^{(1)}(l)/m$, $\tilde\rho=p^\mu\epsilon^{(2)}_\mu(l)/m$. Note that $\mu=q^2-m^2$ and $l^2$ have the meaning of the electron and photon virtualities in the loop respectively. After these substitutions the integrals over $\rho$ and $\nu'$ are trivial, and the remaining 4-dimensional $\delta$-function provides the diagonality of the mass operator in the $E_p$-representation, $M(p',p)=(2\pi)^4\delta^{(4)}(p'-p) M(p)$. This diagonality is expected due to the translational symmetry of the CCF, as $M(p',p)$ is gauge invariant. Even though we sum only a subclass of diagrams, $M(p',p)$ is indeed gauge invariant, as the bubble-chain dressed photon propagator is transverse. Finally, the variable $\tilde\rho$ can be integrated out by employing the formula 
\begin{equation}\label{AiAi2Ai}
\int_{-\infty}^\infty d\tilde\rho \mathrm{Ai}^2(a+\tilde\rho^2)= \frac{1}{2} \mathrm{Ai}_1(2^{2/3} a),
\end{equation}
where 
\begin{equation}\label{Ai1_def}
\mathrm{Ai}_1(t)=\int_t^\infty\,\mathrm{Ai}(x)\,dx=\frac{-i}{2\pi}\int_{-\infty}^\infty \frac{d\sigma}{\sigma-i0} \, e^{-i(t \sigma+\sigma^3/3)}
\end{equation}
is the Aspnes function, see section 3.5.2 and Eq.~(3.105) in \cite{vallee2004airy}. Note that the dimensionless integration variable $\sigma$ in Eq.~\eqref{Ai1_def}, arising after application of Eq.~\eqref{AiAi2Ai}, is proportional to the phase formation interval of the outer loop. After these simplifications, the final expression contains three integrations: over $u$ and the virtualities $l^2$ and $\mu$. In addition, several integrations are 'hidden' in the definition of the Airy functions and in the final form of the bubble-chain dressed photon propagator [see Eqs.~\eqref{photon_prop}, \eqref{photon_prop_trans}, and \eqref{pi12}]. 

After substituting the mass operator into the invariant amplitude $\mathcal{M}(\chi)\equiv \bar{u}_{p,\lambda} M(p) u_{p,\lambda}$, where $u_{p,\lambda}$ is the free Dirac spinor, $p^2=m^2$, and $\lambda$ indicates a spin state, and evaluating the resulting spinor matrix elements, it is natural to split $\mathcal{M}$ into two terms,
\begin{equation}\label{M_split}
\mathcal{M}(\chi)=\mathcal{M}^{(0)}(\chi)+\delta\mathcal{M}(\chi), 
\end{equation}
where 
\begin{widetext}
\begin{eqnarray}
\label{M0}
\begin{aligned}
&\mathcal{M}^{(0)}(\chi)=\frac{\alpha m^2}{2\pi^2} \int_{-\infty}^{\infty} \frac{du}{(1+u)^2}\, \int_{-\infty}^{\infty} d l^2 \int_{-\infty}^{\infty} \frac{d\mu}{\mu+i0}\,D_0(l^2,\chi_l)  \\
&\quad\quad\quad\quad \times \left[{\rm Ai}_1(t) + \frac{u^2+2u+2}{1+u}\left(\frac{\chi}{u}\right)^{2/3}{\rm Ai}'(t) -\frac{2\gamma_s}{(1+u)} \left(\frac{u}{\chi}\right)^{2/3} {\rm Ai}(t)  \right],
\end{aligned}
\end{eqnarray}
\end{widetext}
\begin{eqnarray}
\label{t}
t=\left(\frac{u}{\chi}\right)^{2/3}\left(1+\frac{1+u}{u^2}\frac{l^2}{m^2} +\frac{1+u}{u}\frac{\mu}{m^2}\right),\\
\label{chi_l}
\chi_l=\frac{u\chi }{1+u},
\end{eqnarray}
corresponds to the 1-loop contribution (i.e., it contains no vacuum polarization insertions, see the first diagram in Fig.~\ref{fig:diagram}). This leading-order result has already been calculated and discussed by Ritus \cite{ritus1972radiative}. The (not necessarily small) modifications induced by vacuum polarization are denoted as $\delta\mathcal{M}(\chi)$. Here $\gamma_s=-e F^*_{\mu\nu}p^\mu s^\nu /2m^3$ and $s^\nu=\bar{u}_{p,\lambda} \gamma^\nu\gamma^5 u_{p,\lambda}/2m$ is the electron spin 4-vector \cite{berestetskii1982quantum, meuren2015nonlinear}.

Note that the mass operator needs to be renormalized before physically meaningful quantities can be inferred. According to the standard procedure, this is done successively by proceeding from inner to outer loops. However, if one employs the renormalized polarization operator from the beginning, only the outer (photon) loop remains to be renormalized. This is achieved by adding and subtracting the field-free amplitude $\mathcal{M}(F=0)$, which is renormalized in the standard way and vanishes on-shell \cite{ritus1972radiative}. In case of $\mathcal{M}^{(0)}(\chi)$ this implies that we have to replace the function $\mathrm{Ai}_1(t)$ in Eq. \eqref{M0} with
\begin{equation}
\mathrm{Ai}^{\mathrm{(ren)}}_1(t)=\frac{-i}{2\pi}\int_{-\infty}^\infty \frac{d\sigma}{\sigma} \, e^{-it \sigma}\left(e^{-i\sigma^3/3}-1\right).
\end{equation}
In the following we assume this replacement in $\mathcal{M}^{(0)}(\chi)$ by default without explicitly changing our notation. After renormalization, $\mathcal{M}^{(0)}$ exhibits the following asymptotic scaling for $\chi\gg 1$ [see Eq.~(72) in Ref.~\cite{ritus1972radiative} and Table~\ref{tab:known}, diagram (1b)]
\begin{equation}
\label{M0_asympt}
\begin{split}
{\cal M}^{(0)}(\chi\gg 1)&\approx e^{-i\pi/3}\, \frac{28\sqrt[6]{3}}{27}\Gamma\left(\frac{2}{3}\right) \alpha\chi^{2/3}m^2\\
&\simeq 0.843 (1-i\sqrt{3}) \alpha\chi^{2/3}m^2,
\end{split}
\end{equation}
where $\Gamma(\zeta)$ is the Euler $\Gamma$-function.

The non-trivial contribution $\delta\mathcal{M}(\chi)=\delta\mathcal{M}_1(\chi)+\delta\mathcal{M}_2(\chi)$ in Eq.~\eqref{M_split} is given by
\begin{widetext}
\begin{equation}
\label{deltaM}
\begin{split}
\delta{\cal M}_{1,2}(\chi)=&-\frac{\alpha m^2}{(2\pi)^2}\int_{-\infty}^{\infty}\frac{du}{(1+u)^2}\,\int_{-\infty}^{\infty} dl^2 \int_{-\infty}^{\infty} \frac{d\mu}{\mu+i0}\,   D_{1,2}(l^2,\chi_l)\\
&\times\left[ \vphantom{\left(\frac{u}{\chi}\right)^{2/3}} \left( 1+\frac{l^2}{m^2}\frac{u^2+2u+2}{2u^2}\right){\rm Ai}_1(t) + \left(\frac{u^2+2u+2}{1+u}\pm 1\right)\left(\frac{\chi}{u}\right)^{2/3}{\rm Ai}'(t) -2\gamma_s\left(\frac{1}{1+u}\pm 1 \right) \left(\frac{u}{\chi}\right)^{2/3} {\rm Ai}(t)  \right],
\end{split}
\end{equation}
\end{widetext}
where $t$ and $\chi_l$ are defined in Eqs.~\eqref{t} and \eqref{chi_l}, respectively. Unlike $\mathcal{M}^{(0)}$, these terms vanish as the field is switched off, hence they remain unaffected by renormalization. Apart from the spin-dependent terms, which we write here explicitly, our expression in Eq.~\eqref{deltaM} is equivalent to Eq. (42) in \cite{narozhny1980expansion}, where the factor $Z$ was also set to unity\footnote{According to our investigation the extra overall factor $1/|\chi|$ present in \cite{narozhny1980expansion} is a typo. As we discuss in the next section, the extra terms proportional to $\mu$ inside the coefficients of the Airy and Aspnes functions in Eq. (42) of \cite{narozhny1980expansion} actually vanish after integration.}. 

So far we have mainly followed \cite{ritus1972radiative}. From now on, however, we will proceed differently than in the existing literature \cite{ritus1972radiative,morozov1975elastic,morozov1977elastic,narozhny1979radiation,narozhny1980expansion},
which now applied a perturbative expansion
\begin{equation}\label{D_expansion}
\frac{\Pi_{1,2}(l^2,\chi_l)}{l^2-\Pi_{1,2}(l^2,\chi_l)}=\sum\limits_{r=1}^\infty \left[\frac{\Pi_{1,2}(l^2,\chi_l)}{l^2+i0}\right]^{r},
\end{equation}
in Eq.~\eqref{photon_prop_trans}, where the $r$-th term corresponds to a diagram with $r+1$ loops, including $r$ vacuum polarization bubbles (see Fig.~\ref{fig:diagram}). Here, after reviewing and generalizing this approach, we carry out a non-perturbative calculation and derive the large-$\chi$ asymptotic scaling of the whole amplitude given in Eq.~\eqref{deltaM}. In order to achieve these goals, we process the outer integrals in a different order than in Refs.~\cite{ritus1972radiative,narozhny1980expansion}.

\section{Analysis and all-order resummation of the bubble-type radiative corrections}
\label{sec:virt_int}

\subsection{All-order perturbative analysis}
\label{subsec:pert}

Previous derivations \cite{ritus1972radiative,narozhny1980expansion} of the elastic scattering amplitude were based on a perturbative truncation of the expansion in Eq.~\eqref{D_expansion} for $r\le 2$.  This approach provides some qualitative insights into the scaling of each order of perturbation theory at $1\ll \chi \lesssim \alpha^{-3/2}$. 

The $r$th term of the $D_{1,2}$ expansion given in Eq.~\eqref{D_expansion} corresponds ro $r$ polarization loop insertions. In order to identify the leading-order scaling for such contributions to Eq.~\eqref{deltaM}, we estimate the order of magnitude of each term. The expression under the integral over $u$ rapidly falls off for $u\gtrsim1$ (i.e. if $\chi_l$ of the photon exceeds $\chi_q$ of the electron in the outer loop), hence for the sake of an order-of-magnitude estimate we can restrict integration to $u\lesssim1$ and drop $u$ in the integrand where possible. As we will see shortly, the effective values of $u$ can be small. Therefore, we retain the dependence on $u$ in all factors blowing up at $u\to 0$.
This, together with Eq.~\eqref{Ai_def}, allows us to approximate (up to a complex numerical coefficient) the term containing $\mathrm{Ai}'(t)$ in Eq.~\eqref{deltaM} as 
\begin{equation}\label{M_est}
\begin{split}
\mathcal{M}_{\mathrm{Ai}'}^{(r)}\sim &\, \alpha m^2 \int du  \int d\sigma\,\sigma\,\left(\frac{\chi}{u}\right)^{2/3}
e^{-i\sigma(u/\chi)^{2/3}-i\sigma^3/3}\\ &\times
\int \frac{dl^2\, e^{-il^2\tau}}{l^2+i0}\left[ \frac{\Pi(l^2,\chi_l)}{l^2+i0}\right]^{r}\int \frac{d\mu\,e^{-i\mu s}}{\mu+i0},
\end{split}
\end{equation}
where $\chi_l\simeq u\chi$ [see Eq.~\eqref{chi_l}], $\Pi(l^2,\chi_l)$ is either $\Pi_1$ or $\Pi_2$ [see Eq.~\eqref{pi12}]. Futhermore,
\begin{gather}
\label{prop_time_electron}
s=s(\sigma,u)=\frac{\sigma}{m^2} \left(\frac{u}{\chi}\right)^{2/3} \frac{(1+u)}{u},\\
\label{prop_time_photon}
\tau=\tau(\sigma,u)=\frac{\sigma}{m^2} \left(\frac{u}{\chi}\right)^{2/3} \frac{(1+u)}{u^2}= \frac{s}{u},
\end{gather}
have dimension of inverse mass squared and are proportional to the proper times of the electron and photon in the outer loop, respectively. The meaning of Eqs.~\eqref{prop_time_electron} and \eqref{prop_time_photon} is that both $s$ and $\tau$ are proportional to the phase formation interval $\sigma$ of the outer loop. Note that even though Eq.~\eqref{M0} differs from \eqref{deltaM}, in our approximation the structure of its term containing $\mathrm{Ai}'(t)$ is the same as in Eq.~\eqref{M_est} with $r=0$, hence we consider $r\ge 0$ in what follows.

By applying a dimension-based argument [namely, by assuming $\mu,\,d\mu\sim \mu_{\text{eff}}$ and $l^2, dl^2 \sim (l^2)_{\text{eff}}$] the integrals over the virtualities $\mu$ and $l^2$ are estimated by $\int d\mu\,e^{-i\mu s}/(\mu+i0)\sim 1$ and
\begin{equation}
\label{int_l2_est}
\int \frac{dl^2}{l^2+i0}\left[ \frac{\Pi(l^2,\chi_l)}{l^2+i0}\right]^{r} e^{-il^2\tau}\sim \left[\frac{\Pi\boldsymbol{(}(l^2)_{\text{eff}},\chi_l\boldsymbol{)}}{(l^2)_{\text{eff}}}\right]^{r},
\end{equation}
where the effective scales of the virtualities are established by 
\begin{gather}
\label{mu_eff}
\mu_{\text{eff}}=\frac1{s}\simeq \frac{m^2\chi^{2/3}u^{1/3}}{\sigma},\\
\label{l2_eff}
(l^2)_{\text{eff}}=\frac1{\tau}\simeq \frac{m^2\chi^{2/3}u^{4/3}}{\sigma}.
\end{gather}

As explained above, the integration range over $u$ is effectively restricted from above by $u\lesssim1$. Similarly, the $\sigma^3$ term in the exponential can be effectively replaced with imposing the restriction $\sigma\lesssim1$. With this, the remaining term in the exponential $\sim\sigma(u/\chi)^{2/3}=\mathcal{O}(\chi^{-2/3})\ll1$ and can be neglected. The restrictions of the remaining integration variables $u$ and $\sigma$ from below follow from the fall-off of $\Pi_{1,2}(l^2,\chi_l)$ for $\chi_l\sim u\chi\lesssim 1$ and for $l^2\gtrsim m^2\chi_l^{2/3}$ (see Figs.~\ref{fig:polop_chi} and \ref{fig:polop_l2} in the Appendix, respectively). Note that $\Pi_{1,2}(l^2,\chi_l)$, as a function of $l^2$, decays exponentially to the left of the origin and exhibits a power law decay at the same scale as it oscillates to the right. Therefore we effectively have $u\gtrsim 1/\chi$ and, in virtue of Eq.~\eqref{l2_eff}, $\sigma\gtrsim u^{2/3}$. Inside this range, we can estimate $\Pi(l^2,\chi_l)\simeq \alpha m^2\chi_l^{2/3}$ [see Eq.~\eqref{polop_asympt_off_shell} in the Appendix] and hence [see Eq.~\eqref{l2_eff}]
\begin{equation}\label{g_meaning}
\frac{\Pi\boldsymbol{(}(l^2)_{\text{eff}},\chi_l\boldsymbol{)}}{(l^2)_{\text{eff}}}\sim \frac{\alpha\sigma}{u^{2/3}}.
\end{equation}

By substituting Eq.~\eqref{g_meaning} into Eq.~\eqref{int_l2_est} and the latter into \eqref{M_est}, we obtain
\begin{equation}\label{M_r_estimate_gen}
\mathcal{M}_{\mathrm{Ai}'}^{(r)}\sim  \alpha^{r+1} m^2\chi^{2/3} \int_{\chi^{-1}}^1 \frac{du}{u^{2(r+1)/3}}\int_{u^{2/3}}^1 d\sigma\,\sigma^{r+1}.
\end{equation}
Here, for any $r\ge 0$, the integral over $\sigma$ is $\sim1$, being formed at $\sigma\sim\sigma_{\text{eff}}=1$. However, the integral over $u$ behaves differently for $r=0$ and $r\geq1$.

For $r=0$ (no bubbles) the value of the integral over $u$ in Eq.~\eqref{M_r_estimate_gen} is formed at $u\sim u_{\text{eff}}=1$. Thus, assuming $du\sim u_{\text{eff}}$, we obtain $\mathcal{M}^{(0)}\sim m^2 g$ and a loop formation scale $m\tau_{\text{eff}}\sim 1/(m\chi^{2/3})$, which are in agreement with Eq.~\eqref{M0_asympt} and \cite{yakimenko2019prospect}.

In contrast, for $r\geq1$, $u$ shows up in the denominator of the integrand in Eq.~\eqref{M_r_estimate_gen} in the power $2(1+r)/3\ge1$. This means that the integrand rapidly falls off on this scale and the integral in $u$ is actually formed around the lower limit $u\sim 1/\chi\ll1$. Therefore, we obtain\footnote{Though our reasoning is almost similar to the one given in Ref.~\cite{narozhny1980expansion}, we emphasize several important aspects which are missing there, in particular regarding the composition of the parameter $g$ and the origin of the overall suppression of higher orders in elastic scattering.}
\begin{equation}\label{M_r_estimate}
\mathcal{M}_{\mathrm{Ai}'}^{(r\geq 1)}\sim  m^2 \frac{g^{r+1}}{\chi^{1/3}},\quad g=\alpha\chi^{2/3}.
\end{equation}
This clarifies that for $r\ge1$ the effective value of the photon virtuality is small, $(l^2)_{\text{eff}}/m^2\sim 1/\chi^{2/3}\ll 1$, and the loop formation scale is different, $m\tau_{\text{eff}}\sim \chi^{2/3}/m$.

So far we have only considered the terms $\propto \mathrm{Ai}'(t)$. Let us now discuss the other contributions in Eqs.~\eqref{M0} and \eqref{deltaM}. Obviously, in both of them the terms containing $\mathrm{Ai}_1(t)$ are estimated the same way as above by Eq.~\eqref{M_est}, with the only replacement $\sigma\,(\chi/u)^{2/3}\mapsto 1/\sigma$ in the preexponential factor of the integrand for Eq.~\eqref{M0} and $\sigma\,(\chi/u)^{2/3}\mapsto l^2/(m^2u^2\sigma)\sim (\chi/u)^{2/3}/\sigma^2$ for Eq.~\eqref{deltaM}. Then it is easy to see that in the case $r=0$, for which as before $\sigma,u\sim1$, no enhancement by powers of $\chi$ occurs, hence $\mathcal{M}_{\mathrm{Ai}_1}^{(0)}$ can be neglected against $\mathcal{M}_{\mathrm{Ai}'}^{(0)}$ for $\chi\gg1$. For $r\ge1$ we obtain, instead of Eq.~\eqref{M_r_estimate_gen},
\begin{equation}\label{M_r_estimate_gen1}
\mathcal{M}_{\mathrm{Ai}_1}^{(r\ge1)}\sim  \alpha^{r+1} m^2\chi^{2/3} \int_{\chi^{-1}}^1 \frac{du}{u^{2(r+1)/3}}\int_{u^{2/3}}^1 d\sigma\,\sigma^{r-2}.
\end{equation}
For $r\ge2$ the estimates follow the same derivation as in Eq.~\eqref{M_r_estimate_gen} and the scaling agrees with Eq.~\eqref{M_r_estimate}. However, for $r=1$, the integral over $\sigma$ is formed at small $\sigma\sim\chi^{-2/3}$, which results in an additional factor $\log{\chi}$ [cf. (2b) in Table~\ref{tab:known}]. 

In fact, the calculation to this order was accurately considered in Ref.~\cite{ritus1972radiative}. For $\chi\gg1$ the result is given by Eq.~(76) therein, which, in our notation and up to the accuracy we adopt, can be represented as
\begin{equation}
\label{ritus_eq76}
\mathcal{M}^{(1)}\simeq-\frac{13g^2m^2}{18\pi\sqrt{3}\chi^{1/3}}\left[\frac{\pi}2+
i\left(\ln{\frac{\chi}{2\sqrt{3}}}
-C-\frac{142}{39}\right)\right],
\end{equation}
where $C$ is the Euler constant.

Finally, the terms containing ${\mathrm{Ai}}(t)$ in Eq.~\eqref{M0} and \eqref{deltaM} can be estimated by replacing in the integrand of Eq.~\eqref{M_est} $\sigma\,(\chi/u)^{2/3}\mapsto \gamma_s (u/\chi)^{2/3}$. Then it turns out that $\sigma\sim1$ for all $r\ge0$, but $u\sim1$ for $r=0,\,1,\,2$ and $u\sim \chi^{-1}$ for $r\ge 3$. Furthermore, by estimating $\gamma_s\sim\chi$, we obtain, $\mathcal{M}_{\mathrm{Ai}}^{(r)}\simeq \alpha^{r+1}m^2\chi^{1/3}$ for $r\le 2$ and $\mathcal{M}_{\mathrm{Ai}}^{(r)}\simeq m^2g^{r+1}/\chi^2$ for $r\ge3$. This proves that the spin-dependent contributions also get enhanced at higher orders. However, as implied in \cite{ritus1972radiative, narozhny1980expansion}, they still remain subleading at all orders for $\chi\gg1$.

\begin{table*} %[h!]
	\caption{\label{tab:scales}Summary of the scales for the perturbative and resummed bubble-type mass corrections.}
	\renewcommand*{\arraystretch}{2.1}
	\begin{ruledtabular}
		\begin{tabularx}{\textwidth}{lcccccc}
			& \multicolumn{3}{c}{Perturbative ($1\ll\chi\ll\alpha^{-3/2}$)} & \multicolumn{3}{c}{Resummed ($\alpha\chi^{2/3}\gtrsim1$)} \\
			& $\mathcal{M}^{(0)}$ & $\mathcal{M}^{(1)}$ & $\mathcal{M}^{(r\ge 2)}$& $\delta\mathcal{M}^{(\mathrm{I})}$ 
			& $\delta\mathcal{M}^{(\mathrm{II})}$ & $\delta\mathcal{M}^{(\mathrm{III})}$\\
			\hline
			Scaling [$m^2$]  & $\alpha\chi^{2/3}$ & $\alpha^2\chi\log{\chi}$  & $\alpha^{r+1}\chi^{(2r+1)/3}$ & $\alpha^2$& $\alpha^{3/2}\chi^{2/3}$  & $\alpha^2\chi\log(\alpha^{-3/2})$ \\
			Dominant contribution & $\mathcal{M}_{\mathrm{Ai}'}^{(0)}$ & $\mathcal{M}_{\mathrm{Ai}_1}^{(1)}$ & $\mathcal{M}_{\mathrm{Ai}'}^{(r)}\sim \mathcal{M}_{\mathrm{Ai}_1}^{(r)}$& - & - & -\\
			$\sigma$ & 1& $\chi^{-2/3}$& 1   & 1& 1& $(\alpha\chi^{2/3})^{-1}$ \\
			$u$ & 1 & $\chi^{-1}$&  $\chi^{-1}$ & 1 & $\alpha^{3/2}$&  $\chi^{-1}$\\
			$\chi_l\sim u\chi$ \footnotemark[1] & $\chi$ & 1 & 1 & $\chi$ & $\alpha^{3/2}\chi$& 1\\
			$\tau\sim\dfrac{\sigma}{m^2\chi^{2/3}u^{4/3}}$ \footnotemark[2] [$m^{-2}$] & $\chi^{-2/3}$ & 1 & $\chi^{2/3}$ & $\chi^{-2/3}$ & $\alpha^{-2} \chi^{-2/3}$& $\alpha^{-1}$ \\
			$\dfrac{\tau}{\tau_{\text{eff}}^{(1)}}$ & 1 & 1& $\chi^{2/3}$ & 1 & $\alpha^{-1}$ & $\alpha^{-1}$ \\
			$l^2\sim \tau^{-1}$ [$m^2$] & $\chi^{2/3}$ & 1 & $\chi^{-2/3} $ & $\chi^{2/3}$ & $\alpha^2 \chi^{2/3}$ & $\alpha $ \\
			$s=u\tau $ \footnotemark[3]  [$m^{-2}$]&  $\chi^{-2/3}$ & $\chi^{-1}$ & $\chi^{-1/3}$ &  $\chi^{-2/3}$ & $\alpha^{-1/2} \chi^{-2/3}$ & $\alpha^{-1}\chi^{-1}$ \\
			$\mu\sim s^{-1}$ [$m^2$] & $\chi^{2/3}$ & $\chi$ & $\chi^{1/3}$ & $\chi^{2/3}$ & $\alpha^{1/2}\chi^{2/3}$ & $\alpha \chi $ \\
			$\dfrac{\Pi(l^2,\chi_l)}{l^2}\sim \dfrac{\alpha\sigma}{u^{2/3}}$ \footnotemark[4] & $\alpha$ & $\alpha$ & $\alpha\chi^{2/3}$ & $\alpha$ & 1 & 1\\
	\end{tabularx}
	\end{ruledtabular}
	\footnotetext[1]{See Eq.~\eqref{chi_l}.}
	\footnotetext[2]{See Eq.~\eqref{prop_time_photon}.}
	%\footnotetext[3]{Note that $\tau_{\rm eff}^{(1)}=1/(m^2\chi_l^{2/3})$.}
	\footnotetext[3]{See Eq.~\eqref{prop_time_electron}.}
	\footnotetext[4]{See Eq.~\eqref{g_meaning}.}
\end{table*}

To summarize, we have reproduced the asymptotic scalings of the diagrams (1b), (2b) and (3g) in Table~\ref{tab:known}. Moreover, the above analysis extends these results to all orders, thereby establishing this aspect of the Ritus-Narozhny conjecture. The findings of this section for the scales of the leading-order perturbative contributions are collected in the first three columns of Table~\ref{tab:scales}. In the following we will compare them to the scaling naturally arising after the all-order resummation. 

Before proceeding, however, we would like to point out a few important insights. The scalings of the corrections at all orders (apart from the $\log{\chi}$-factor occurring solely for $r=1$ as discussed above) are consistent with a direct estimate $\mathcal{M}^{(r)}\simeq \alpha  s^{-1} (\Pi/l^2)^r$ based on Fig.~\ref{fig:diagram}, where the factors $\Pi$ and $\alpha$ come from each bubble and the two remaining vertices, $s^{-1}$ and $l^{-2}$ correspond the electron and photon propagators, respectively (with the specific appropriate choice of all the scales for given $r\ge0$). Hence it is clear that for $r\ge 2$ the scaling parameter $g$ naturally originates as the ratio of the polarization operator eigenvalues to the characteristic value of the photon virtuality in Eq.~\eqref{g_meaning}. However, in the special lowest-order cases $r=0$ and $r=1$ this ratio acquires standard for field-free QED value $\alpha$. This, however, is accompanied with a variation of the loop formation scales for $r\le 2$, which become uniform only for $r\ge2$. The latter includes a modification of either the characteristic values of $u$ (equivalently, $\chi_l$) or $\sigma$, or even an alteration of the dominant contribution, and explains the anomalous ratio $\sim \alpha\chi^{1/3}\log\chi$ of the two-loop and the one-loop mass corrections mentioned in the introduction. In effect, however, the latter ratio becomes uniform already for $r\ge1$, as, disregarding the $\log{\chi}$-factor, the corrections at these orders are all estimated by Eq.~\eqref{M_r_estimate}. As compared to $\mathcal{M}^{(0)}$, the resulting scaling contains an extra factor $\chi^{-1/3}$.

Due to the presence of this extra factor in higher-loop diagrams one has to distinguish between the critical value $g\sim1$, for which all higher-order terms become of the same order and the perturbative expansion breaks down, and the regime $g\gg 1$, where higher-order terms become comparable to the 1-loop contribution $\mathcal{M}^{(0)}$ and thus substantially modify the total amplitude. This was nicely rephrased in \cite{dixon2019potentially}, by observing that for $\chi=\alpha^{-3/2}$ (i.e. $g=1$), the bubble-type corrections (2b) and (3g) in Table \ref{tab:known} are both suppressed with respect to (1b) by the same factor $\sqrt{\alpha}$, whereas for larger values of $\chi$, e.g. for $\chi\sim \alpha^{-2}$, they are growing with $r$ and hence may compete with (1b). As we have shown here, the same happens for the higher-order ($r\ge3$) corrections as well.

\subsection{All-order resummation at $\alpha\chi^{2/3}\gtrsim 1$}
\label{subsec:appr}
After the qualitative discussion of the perturbative scaling in the previous section we now present a quantitative analysis in the non-perturbative regime $\alpha\chi^{2/3}\gtrsim1$. In principle, this could be done by an all-order resummation of the perturbative bubble-type contributions $\mathcal{M}^{(r)}$. Such a procedure, however, is hardly implementable, as the numerical coefficient of $\mathcal{M}^{(r)}$ is defined by nested integrals and their overall number grows substantially at higher orders. Therefore, it is more practical to evaluate Eq.~\eqref{deltaM} directly. In essence, our calculation is fully equivalent to a Borel summation \cite{zinn1981perturbation} of the all-order bubble-type diagrams in Fig.~\ref{fig:diagram}.

We proceed by employing the integral representations for the Airy \eqref{Ai_def} and the Aspnes function \eqref{Ai1_def}, and changing the order of integration by considering first the integrals over the virtualities $\mu$ and $l^2$. Then the integral over $\mu$ reduces to the textbook form
\begin{equation}\label{int_mu}
\int_{-\infty}^{\infty} d\mu\, \frac{e^{-i\mu s}}{\mu+i0}=-2\pi i\,\theta\left(\mathrm{Re}\,s\right),
\end{equation}
where $\theta$ is the Heaviside step function. Here we treat the parameter $s$  complex-valued if $u$ is negative. Note that any contribution to the coefficients of the Airy and Aspnes functions in Eq.~\eqref{deltaM}, which is linear in $\mu$ (cf. Ref.~\cite{narozhny1980expansion}), vanishes at this point. To show this we note that
\begin{equation}\label{int_mu1}
\int_{-\infty}^{\infty} d\mu\, \frac{\mu \,e^{-i\mu s}}{\mu+i0}=2\pi \delta(s).
\end{equation}
Hence such terms do not contribute after the integration over $\sigma$ is carried out (more details are given below).

Next we consider the integral over $l^2$, which is more involved, but can be suitably approximated. After substituting Eq.~\eqref{photon_prop_trans} into Eq.~\eqref{deltaM} we obtain two kinds of integrals\footnote{Note that the subscripts $1,2$ correspond to the two different values given in curly brackets.}
\begin{equation}
\label{J12}
\begin{split}
J^{(i)}_{1,2}(\tau,\chi_l)=\int_{-\infty}^{\infty} dl^2\, &\left\{\frac{l^2}{m^2}\,, 1 \right\}\,\\
&\times\frac{\Pi_i(l^2,\chi_l) e^{-i l^2 \tau}}{(l^2+i0)[l^2-\Pi_i(l^2,\chi_l)]},
\end{split}
\end{equation}
where $\Pi_i(l^2,\chi_l)$ is either $\Pi_1$ or $\Pi_2$ [see Eq.~\eqref{pi12}]. Note that the components of the polarization operator admit a one-sided Fourier integral representation
\begin{equation}\label{pi_fourier}
\Pi_i(l^2,\chi_l)=\int_0^\infty d\tau\,\tilde{\Pi}_i(\tau,\chi_l)\,e^{il^2\tau},
\end{equation}
where $\tilde{\Pi}_{1,2}(l^2,\chi_l)$ are given in Eq.~\eqref{pi_Fourier}. We combine Eq.~\eqref{pi_fourier} with the complete perturbative expansion given in Eq.~\eqref{D_expansion} and rewrite $J^{(i)}_1$ as
\begin{equation}\label{J_exact}
\begin{split}
J^{(i)}_1(\tau,\chi_l)=&-\frac{2\pi i}{m^2}\sum_{n=0}^\infty\frac{(-i)^n}{n!}\left[\prod_{a=1}^{n+1}\int\limits_0^\infty d\tau_a\,\tilde{\Pi}_i(\tau_a,\chi_l)\right]\\
&\times\left(\tau-\sum_{a=1}^{n+1}\tau_a\right)^n \theta\,\left(\mathrm{Re}\,\tau-\sum_{a=1}^{n+1}\tau_a\right).
\end{split}
\end{equation}
Here and below, unless stated otherwise, we consider explicitly only $J^{(i)}_1(\tau,\chi_l)$, implying that $J^{(i)}_2(\tau,\chi_l)$ is handled in the same way. 

\begin{figure}
	\begin{center}
		\includegraphics[width=\columnwidth]{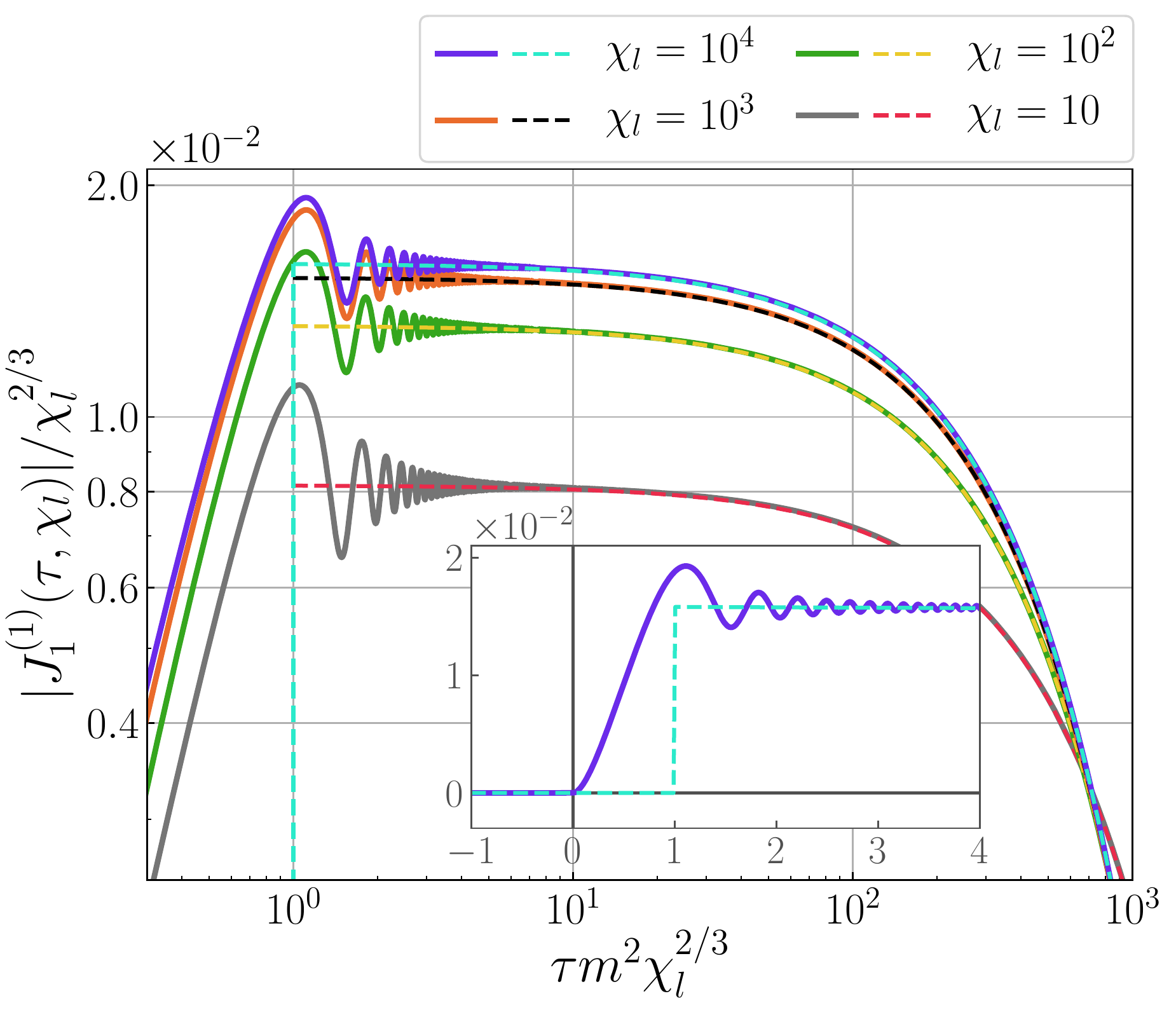}
	\end{center}
	\caption{A test of the approximation given in Eq.~\eqref{J1_approx} for $J^{(1)}_1(\tau,\chi_l)$ (dashed lines) against its direct numerical evaluation (solid lines) shown in a double-logarithmic scale for $\chi_l=10,\,10^2,\,10^3\,\text{and}\,10^4$ (the inset shows the same in a linear scale for $\chi_l=10^4$). The dashed vertical line corresponds to the value $\tau=\tau_{\rm eff}^{(1)}$.}
	\label{fig:J1}
\end{figure}

In the following we mainly focus on the asymptotic region $\chi\gtrsim\alpha^{-3/2}$ ($g\gtrsim1$) and derive an approximation which is valid in this regime. As we will see further, the effective value of $\chi_l$, that corresponds to the dressed photon, does not necessarily obey the same condition, yet $\chi_l\gtrsim 1$. For $\chi_l\gtrsim 1$ the value of the integrals over $\tau_a$ are effectively accumulated at $\tau_a\lesssim\tau_{\rm eff}^{(1)}=1/(m^2\chi_l^{2/3})$.  

Next, we use an ad hoc approximation, which we will substantiate below: we neglect $\tau_a$ compared to $\tau$ in the second line of Eq.~\eqref{J_exact}. Then we obtain
\begin{align}\label{J_exact1}
J^{(i)}_1(\tau,\chi_l)\approx-\frac{2\pi i}{m^2}\theta\left(\mathrm{Re}\,\tau\right)\sum_{n=0}^\infty\frac{(-i)^n}{n!}\Pi_i^{n+1}(0,\chi_l)\tau^n ,
\\ \nonumber
 \Pi_i(0,\chi_l)=\int_0^\infty d\tau_a \tilde{\Pi}_i(\tau_a,\chi_l).
\end{align}
This implies that we can further resum the series to an exponential. For $\chi_l\gtrsim 1$ this simplification is formalized by the observation that for $\chi\gg1$ the values of $\tau$ that effectively contribute in all higher ($r\ge2$) order perturbative contributions are much larger than $\tau_{\mathrm{eff}}^{(1)}$, see Table~\ref{tab:scales}. The same is true also after resummation, as in this case the contribution to the outer integrals is dominated by [see Eq.~\eqref{polop_asympt}]
\begin{equation}\label{z_char}
\tau\sim \tau_{\mathrm{eff}}=\Pi_i^{-1}(0,\chi_l)\sim\tau_{\mathrm{eff}}^{(1)}/\alpha \gg \tau_{\mathrm{eff}}^{(1)}.
%\sim\left(\alpha\chi_l^{2/3}m^2\right)^{-1}.
\end{equation}
However, we have to be careful and should in addition ensure that $J^{(i)}_{1,2}(\tau,\chi_l)$ vanish at $\tau\to 0$, which can be seen from Eq.~\eqref{J_exact}. This property is important, otherwise we would introduce an artificial divergence in the integral over $\sigma$ in the term containing $\mathrm{Ai}_1(t)$. Motivated by this reasoning we come to the following approximation
\begin{equation}\label{J1_approx}
J^{(i)}_1(\tau,\chi_l)\approx -2\pi i\, \theta\,\left(\mathrm{Re}\,\tau-\tau_{\rm eff}^{(1)}\right) \frac{\Pi_i(0,\chi_l)}{m^2} e^{-i\Pi_i(0,\chi_l)\tau}.
\end{equation}
Furthermore, we write
\begin{equation}\label{J2_approx}
J^{(i)}_2(\tau,\chi_l)\approx -2\pi i\, \theta\,\left(\mathrm{Re}\,\tau-\tau_{\rm eff}^{(1)}\right) \left[e^{-i\Pi_i(0,\chi_l) \tau}-1\right],
\end{equation}
where, unlike for $J^{(i)}_1$, the insertion of the $\theta$-function is no more mandatory. For the sake of uniformity, however, we include it also for $J^{(i)}_2$, as the modification doesn't change the asymptotic limit $\chi \to \infty$.

The approximations given in Eqs.~\eqref{J1_approx} and \eqref{J2_approx} are crucial for the analytical derivation of the non-perturbative asymptotic expansion. Therefore, we have verified their validity numerically by comparing Eq.~\eqref{J1_approx} with an exact evaluation of $J^{(1)}_1(\tau,\chi_l)$ based on the definition [see Eq.~\eqref{J12}]. The result is shown in Fig.~\ref{fig:J1}, where we scaled the axes such that the graph converges in the limit $\chi_l\to\infty$. The numerical calculation clearly demonstrates that the approximation given in Eq.~\eqref{J1_approx} is in excellent quantitative agreement with the exact expression for $\tau\gg \tau_{\rm eff}^{(1)}$. Moreover, it ensures, due to the insertion of the Heaviside step function, that $J^{(1)}_1(\tau,\chi_l)$ vanishes at $\tau\to 0$. Finally, we would like to point out that the graph in Fig.~\ref{fig:J1} has a log-log scale. Therefore, the region $\tau\lesssim\tau_{\rm eff}^{(1)}$, where the approximation is poor, doesn't contribute significantly to a well-behaved integral over the full range of $\tau$. 
%Finally, as can be seen from Fig.~\eqref{fig:J1}, the approximation \eqref{J1_approx} appears to be accurate even for $1\ll \chi_l\ll \alpha^{-3/2}$. The whole curve shifts upwards as $\chi_l$ is increased and stabilizes for $\chi_l\gtrsim\alpha^{-3/2}$.
 %it worth noting that the plot in Fig.~\ref{fig:J1} captures the behavior of the function $J_1(\tau,\chi_l)$ for a wide range of $\chi_l$ including $1\ll \chi_l\ll \alpha^{-3/2}$, though we did not demand this when introduced our approximation. E.g. with the change of $\chi_l$ by $-3$ orders of magnitude the whole graph shifts down by a factor of $\sim1$, while the accuracy of the approximation \eqref{J1_approx} remains on the same level as in the example.

After evaluating the integrals over $d\mu$ and $dl^2$ one encounters the following product of Heaviside step functions [see Eqs.~\eqref{int_mu}, \eqref{J1_approx}, and \eqref{J2_approx}], which can be transformed into
\begin{equation}
\begin{aligned}
\theta\left(\mathrm{Re}\,s\right) \theta\left(\mathrm{Re}\,\tau-\tau_{\rm eff}^{(1)}\right)=\theta(u)\theta\boldsymbol{(}\sigma-\sigma_0(u)\boldsymbol{)},\\
\sigma_0(u)=\left[u^2/(1+u)\right]^{1/3}.
\end{aligned}
\end{equation} 

Finally, after applying the derived approximations to Eq.~\eqref{deltaM}, we obtain the resummed amplitude $\delta\mathcal{M}$ valid at $\chi\gg1$. It is convenient to split it into three parts:
\begin{equation}
\label{deltaM_double}
\delta{\cal M}_{i}(\chi)=\delta{\cal M}^{(\mathrm{I})}_i(\chi)+\delta{\cal M}^{(\mathrm{II})}_i(\chi)+\delta{\cal M}^{(\mathrm{III})}_i(\chi),
\end{equation}
where $ i=1,2$ and
\begin{widetext}
\begin{align}
\label{deltaM_I}
\delta{\cal M}^{(\mathrm{I})}_{1,2}(\chi)&=\frac{\alpha m^2}{2\pi}\int_0^\infty \frac{du}{(1+u)^2}\int_{\sigma_0(u)}^\infty \frac{d\sigma}{\sigma}\,e^{-i\sigma^3/3-i\sigma (u/\chi)^{2/3}} \left[e^{-i g\sigma\varphi_{1,2}(u)}-1\right],\\
\label{deltaM_II}
\begin{split}
\delta{\cal M}^{(\mathrm{II})}_{1,2}(\chi)&=\frac{\alpha m^2}{2\pi}\int_0^\infty \frac{du}{(1+u)^2}\left(\frac{\chi}{u}\right)^{2/3}\left(\frac{u^2+2u+2}{1+u}\pm 1\right) \int_{\sigma_0(u)}^\infty d\sigma\,\sigma\,e^{-i\sigma^3/3-i\sigma (u/\chi)^{2/3}}  \left[e^{-i g\sigma\varphi_{1,2}(u)}-1\right],
\end{split}\\
\label{deltaM_III}
\begin{split}
\delta{\cal M}^{(\mathrm{III})}_{1,2}(\chi)&=\frac{\alpha g m^2}{4\pi}\int_0^\infty \frac{du}{(1+u)^2}\left(\frac{\chi}{u}\right)^{2/3}\frac{u^2+2u+2}{1+u}\varphi_{1,2}(u)\int_{\sigma_0(u)}^\infty \frac{d\sigma}{\sigma}\,e^{-i\sigma^3/3-i\sigma (u/\chi)^{2/3}} e^{-i g\sigma\varphi_{1,2}(u)},
\end{split}
\end{align}
\end{widetext}
where $\delta{\cal M}^{(\mathrm{I})}$ and $\delta{\cal M}^{(\mathrm{III})}$ originate in the term initially containing $\mathrm{Ai}_1(t)$ and $\delta{\cal M}^{(\mathrm{II})}$ in the term containing $\mathrm{Ai}'(t)$. Here we introduced the abbreviations $\varphi_i(u)=(1+u)\pi_i(\chi_l)/(\chi u)^{4/3}$ and $\pi_i(\chi_l)=\Pi_i(l^2=0,\chi_l)/(\alpha m^2)$. Notably, at this stage the on-shell eigenvalues of the polarization operator in a CCF $\Pi_{1,2}(l^2=0,\chi_l)$ are exponentiated. As we show below, this results in a modification of the formation scales and asymptotic behavior of the contributions $\delta{\cal M}^{(\mathrm{II,\,III})}$ (though not of $\delta{\cal M}^{(\mathrm{I})}$) in the non-perturbative regime $\alpha\chi^{2/3}\gtrsim 1$.
%Notably, the exponential dependence on the on-shell eigenvalues of the polarization operator in a CCF $\Pi_{1,2}(l^2=0,\chi_l)$ manifests the non-perturbative nature of this result. 

\section{Asymptotic behavior of $\mathcal{\delta M}$ for $\alpha\chi^{2/3}\gg 1$}
\label{sec:u_sigma}
Next we determine the high-$\chi$ asymptotic behavior of each contribution to $\delta\mathcal{M}$ given in Eqs.~\eqref{deltaM_I}, \eqref{deltaM_II} and \eqref{deltaM_III}. It turns out that they exhibit different formation regions, which implies that each contribution also has a different physical interpretation.

\subsection{Contribution $\delta{\cal M}^{(\mathrm{I})}$}
\label{sec:dMI}
In Eq.~\eqref{deltaM_I} it is convenient to change the order of integration in the following way
\begin{equation}\label{swap_usigma}
\begin{aligned}
\int_0^\infty du\int_{\sigma_0}^\infty d\sigma\,\ldots=\int_0^\infty d\sigma\int_0^{u_0(\sigma)} du\,\ldots,\\
 u_0(\sigma)=\frac{\sigma^3}2+\sqrt{\frac{\sigma^6}4+\sigma^3}.
\end{aligned}
\end{equation}
Effectively, the integrals are formed around $\sigma\simeq\sigma_{\rm eff}=1$ and $u\simeq u_{\rm eff}=1$ (to be justified \textit{a posteriori}). This implies that $\chi_l\approx u\chi\sim\chi\gg 1$ [see Eq.~\eqref{chi_l}] and thus $\pi_i(\chi_l)\simeq K_i\chi_l^{2/3}$, where $K_i$ are numerical coefficients defined in the Appendix [see Eq.~\eqref{K_i}]. 

In virtue of the above we can neglect $\sigma(u/\chi)^{2/3}=\mathcal{O}(\chi^{-2/3})$ and retain only the first non-vanishing term of the expansion in the small argument $g\sigma\varphi_{i}(u)\simeq \alpha K_i\ll 1$ of the exponential. Thus, we obtain
\begin{equation}
\label{deltaM_I_approx}
\delta{\cal M}^{(\mathrm{I})}_{i}(\chi)\simeq -C^{(\textrm{I})} K_i \alpha^2 m^2,\quad i=1,2;
\end{equation}
where the coefficient
\begin{equation}\label{C1}
\begin{split}
C^{(\textrm{I})}&=\frac{i}{2\pi}\int_0^\infty d\sigma\,e^{-i\sigma^3/3}\int_0^{u_0(\sigma)} \frac{du}{u^{2/3}(1+u)^{5/3}}\\
&\approx 0.256+0.325i
\end{split}
\end{equation}
is easily evaluated numerically. Note that the formation regions assumed above become transparent in Eq.~\eqref{C1}. 

The resulting contribution $\delta{\cal M}^{(\mathrm{I})}=\sum_{i=1}^2\delta{\cal M}^{(\mathrm{I})}_i=\mathcal{O}(\alpha^2)$ contains no enhancement for $\chi\gg1$. Notably, the expansion of the exponential in $g\sigma\varphi_{i}(u)$, which we showed can be truncated in this case, coincides with a perturbative expansion, and the final contribution to $\delta \mathcal{M}$ is subleading. This is also confirmed by an inspection of the scales of $\delta{\cal M}^{(\mathrm{I})}$ (see Table~\ref{tab:scales}), which coincide with the scales of the leading-order perturbative contribution $\delta{\cal M}^{(0)}$. 

As we will discuss below, there are good reasons in favor of only two physically different non-perturbative contributions. From this perspective $\delta{\cal M}^{(\mathrm{I})}$ should be combined and considered jointly with $\delta{\cal M}^{(\mathrm{III})}$.
This is confirmed by the fact that, unlike $\delta{\cal M}^{(\mathrm{II,III})}$, $\delta{\cal M}^{(\mathrm{I})}$ cannot be estimated as $\alpha/s$. It is also worth pointing out that $\tau_{\rm eff}\simeq 1/\left(m^2\chi_l^{2/3}\right)$ implies that the approximations in Eqs.~\eqref{J1_approx}, \eqref{J2_approx} are actually not sufficient for an accurate calculation of $\delta{\cal M}^{(\mathrm{I})}$. However, as we have shown, $\delta{\cal M}^{(\mathrm{I})}$ is sub-dominant, therefore we do not investigate it any further.

\subsection{Contribution $\delta{\cal M}^{(\mathrm{II})}$}
\label{sec:dMII}

Next we consider Eq.~\eqref{deltaM_II}. It is again convenient to interchange the order of integration using Eq.~\eqref{swap_usigma}. This time, the resulting integral is formed around $\sigma\simeq\sigma_{\rm eff}=1$, but in contrast to $\delta{\cal M}^{(\mathrm{I})}$, around the smaller value $u\simeq u_{\rm eff}= \alpha^{3/2}\ll 1$ (cf. the discussion in Section~\ref{subsec:pert}). Assuming $g=\alpha\chi^{2/3}\gg1$, this still implies $\chi_l\approx u\chi\simeq g^{3/2}\gg 1$ [see Eq. \eqref{chi_l}] and thus, as for $\delta{\cal M}^{(\mathrm{I})}$, $\pi_i(\chi_l)\simeq K_i\chi_l^{2/3}$. The approximations given in Eqs.~\eqref{J1_approx}, \eqref{J2_approx} are valid, since
\begin{equation}\label{tau_eff_II}
\tau_{\mathrm{eff}}\simeq \tau(\sigma_{\mathrm{eff}},u_{\mathrm{eff}}) 
\simeq \frac1{\alpha g m^2}\gg \tau_{\rm eff}^{(1)}, 
\end{equation}
where $\tau(\sigma,u)$ is defined in Eq.~\eqref{prop_time_photon}. As for $\delta{\cal M}^{(\mathrm{I})}$, it is possible to neglect the term $\sigma(u/\chi)^{2/3}=\mathcal{O}(\alpha\chi^{-2/3})$ in the exponential. 

Furthermore, we neglect $u$ due to $u_{\rm eff}\ll 1$ wherever possible and replace the upper limit of the $du$-integral by infinity,
\begin{equation}\label{MII_inter}
\begin{split}
\delta{\cal M}_{1,2}^{\rm (II)}\approx &\frac{(2\pm 1)\alpha\chi^{2/3} m^2}{2\pi}\int_0^\infty d\sigma\,\sigma e^{-i\sigma^3/3} \\
&\times \int_0^{\infty} \frac{du}{u^{2/3}}\,\left(e^{-i K_{1,2}\alpha\sigma /u^{2/3}}-1\right).
\end{split}
\end{equation}
To simplify this expression even further we note that
%We will further comment on this difference in Section~\ref{sec:summ}.
\begin{align}\label{int1}
\int\limits_0^{\infty} \frac{du}{u^{2/3}}\,\left(e^{-i \zeta/u^{2/3}}-1\right)=3e^{i\frac{5\pi}4}\sqrt{\pi\zeta},\\
\label{int2}
\int_0^\infty d\sigma\,\sigma^{3/2} e^{-i\sigma^3/3}=e^{-i\frac{5 \pi}{12}} 3^{-\frac16}\Gamma\left(\frac56\right),
\end{align}
where $\zeta=K_{1,2} \alpha\sigma$.
Finally, we obtain
\begin{equation}
\label{deltaM_II_approx}
\begin{split}
\delta{\cal M}^{\rm (II)}&=\sum\limits_{i=1}^2\delta{\cal M}^{\rm (II)}_i\\
&\simeq e^{i\frac{5\pi}6}\frac{3^{5/6}}{2\sqrt{\pi}}\Gamma\left(\frac56\right)\left(3\sqrt{K_{1}}+\sqrt{K_{2}}\right)\,\alpha^{3/2}\chi^{2/3}m^2\\
&\approx (-0.995+1.72 i)\alpha^{3/2}\chi^{2/3}m^2.
\end{split}
\end{equation}
The integrals in Eqs.~\eqref{int1} and \eqref{int2} are obviously formed at the scales $u\simeq \zeta^{3/2}\simeq \alpha^{3/2}$ and $\sigma\simeq 1$.

\begin{figure}
	\begin{center}
		\includegraphics[width=\columnwidth]{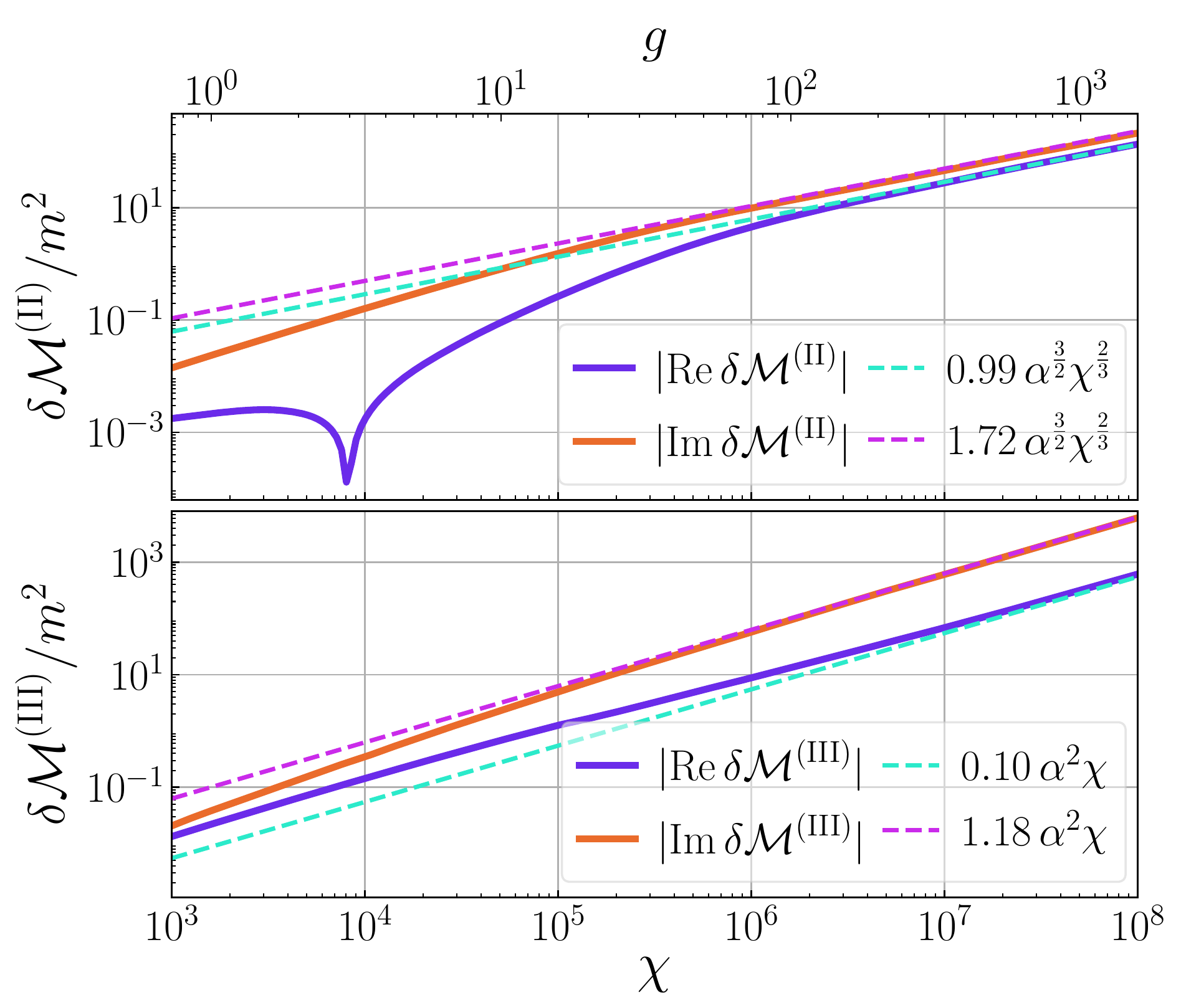}
	\end{center}
	\caption{Dependence of the resummed bubble-type mass correction on $\chi$ (the corresponding value for $g$ is given on the upper axis): asymptotic expressions \eqref{deltaM_II_approx}, \eqref{deltaM_III_approx_final} for $\chi\gg 1$ (dashed lines) vs direct numerical evaluation of \eqref{deltaM_II}, \eqref{deltaM_III} (solid lines).}
	\label{fig:deltaM}
\end{figure}

A numerical comparison between the exact [see Eq.~\eqref{deltaM_II}] and the asymptotic [see Eq.~\eqref{deltaM_II_approx}] expression is shown in the upper panel of Fig.~\ref{fig:deltaM}. One can see that the asymptotics [see Eq.~\eqref{deltaM_II_approx}] is indeed eventually achieved, though for extremely high values $\chi\gtrsim 10^6$ corresponding to $g\sim 100$. Notably, Eq.~\eqref{deltaM_II_approx} overestimates the exact result for smaller $\chi$. The error is particularly large for the real part, which changes sign at $\chi\simeq 8\times 10^3$.

The obtained scales characterizing the correction $\delta{\cal M}^{(\mathrm{II})}$ are listed in Table~\ref{tab:scales}. One can notice that they have the same dependence on $\chi$ as the scales for $\mathcal{M}^{(0)}$, but incorporate the coupling $\alpha$ differently. We will further comment on this difference in Section~\ref{sec:summ}.

\subsection{Contribution $\delta{\cal M}^{(\mathrm{III})}$}
\label{sec:dMIII}

Finally, we consider the last contribution in Eq.~\eqref{deltaM_III}. Here it is convenient to keep the integration order but change the integration variables from $u$ to $\chi_l=u\chi/(1+u)$ and from $\sigma$ to $\tilde{\sigma}=\sigma/\sigma_0(u)$. Assuming $u\simeq u_{\mathrm{eff}}\ll1$ (to be confirmed \textit{a posteriori}) we neglect $u$ where possible, in particular the term $\sigma(u/\chi)^{2/3}$. Thus, we obtain
\begin{equation}\label{deltaM_III_simpl}
\begin{split}
\delta{\cal M}_i^{\rm (III)}\approx & \frac{\alpha^2 m^2\chi}{2\pi}\int_0^\infty \frac{d\chi_l}{\chi_l^2}\,\pi_i(\chi_l)\\
& \times \int_1^\infty \frac{d\tilde{\sigma}}{\tilde{\sigma}}\,e^{-i(\chi_l/\chi)^2\tilde{\sigma}^3/3-i\alpha\tilde{\sigma} \pi_i(\chi_l)/\chi_l^{2/3}}.
\end{split}
\end{equation}
In virtue of $\pi_i(\chi_l\gg1)\simeq K_i\chi_l^{2/3}$, the integrals are effectively truncated from above at $\chi_l\simeq (\chi_l)_{\mathrm{eff}}=1$ and $\tilde{\sigma}\simeq \alpha^{-1}$ for $\alpha\chi^{2/3}\gg 1$. This implies that $u_{\mathrm{eff}}=\chi^{-1}\ll1$ [as initially assumed, cf. the prerequisites to Eq.~\eqref{M_r_estimate}] and $\sigma_{\mathrm{eff}}=(\alpha\chi^{2/3})^{-1}\ll1$. Therefore, our approximations given in Eqs.~\eqref{J1_approx}, \eqref{J2_approx} are justified as 
\begin{equation}\label{M_III_z_check}
\tau_{\rm eff}\simeq \tau(\sigma_{\mathrm{eff}},u_{\mathrm{eff}})\simeq \frac1{\alpha m^2}\gg \tau_{\rm eff}^{(1)}
\end{equation}

Moreover, it is also possible to neglect the first term $\mathcal{O}(g^{-3})$ in the exponential in Eq.~\eqref{deltaM_III_simpl}. As a result, we find that
\begin{equation}
\label{deltaM_III_approx}
\delta{\cal M}^{\rm (III)}_i\simeq C_i^{(\mathrm{III})}\alpha^2\chi m^2,
\end{equation}
where the numerical factors $C_i^{(\mathrm{III})}$ are given by
\begin{equation}
\label{CIII}
\begin{split}
C_{1,2}^{(\mathrm{III})}&=\frac1{2\pi}\int_0^\infty \frac{d\chi_l}{\chi_l^2}\,\pi_{1,2}(\chi_l)\,{\rm E}_1\boldsymbol{\big(}i\alpha\pi_{1,2}(\chi_l)/\chi_l^{2/3}\boldsymbol{\big)}\\
&\approx \begin{cases}-0.0395 - 0.472 i, \\ -0.0634 - 0.703 i.\end{cases}
\end{split}
\end{equation}
Here ${\rm E}_1(\zeta)=\int_1^\infty dt\, e^{-\zeta t}/t$ is  the exponential integral. Correspondingly,
\begin{equation}\label{deltaM_III_approx_final}
\delta{\cal M}^{(\mathrm{III})}=\sum\limits_{i=1}^2 \delta{\cal M}_i^{(\mathrm{III})}=-(0.103+1.18i)\alpha^2\chi m^2.
\end{equation}

A numerical comparison between the asymptotic result in Eq.~\eqref{deltaM_III_approx_final} and the exact expression in Eq.~\eqref{deltaM_III} is shown in the lower panel of Fig.~\ref{fig:deltaM}. Similar as for $\delta{\cal M}^{\rm (II)}$, the asymptotic result becomes reliable for $\chi\sim 10^6$ ($g\sim 100$). However, unlike for $\delta{\cal M}^{\rm (II)}$, it represents a good order-of-magnitude estimate even for smaller $\chi$.

The scales for the correction $\delta{\cal M}^{(\mathrm{III})}$ are collected in the last column of Table~\ref{tab:scales} and depend on $\chi$ mostly in the same way as $\mathcal{M}^{(1)}$, but incorporate the coupling $\alpha$ differently. We observed the same in the previous section by comparing $\delta{\cal M}^{(\mathrm{II})}$ to $\mathcal{M}^{(0)}$. Here, however, the difference in the scalings given in Eqs.~\eqref{deltaM_III_approx_final} and \eqref{ritus_eq76} is less obvious and deserves a more detailed discussion. Both are proportional to $\alpha^2\chi$, but the coefficient in Eq.~\eqref{ritus_eq76} contains $\log{\chi}$, whereas the coefficient in Eq.~\eqref{deltaM_III_approx_final} rather contains $\alpha=g\chi^{-2/3}$ in a quite complicated form, see Eq.~\eqref{CIII}. In particular, at the point $\chi\simeq\alpha^{-3/2}$, we have
\begin{equation}\label{ritus_eq76_bdp}
\mathcal{M}^{(1)}(g\simeq 1)\approx -(0.208+0.255i)\alpha^2\chi m^2,
\end{equation}
which should be compared with Eq.~\eqref{deltaM_III_approx_final}. Furthermore, by approximating $\mathrm{E}_1(\zeta)\approx -\ln{\zeta}-C$ and evaluating the integral over $\chi_l$ in Eq.~\eqref{CIII}, with accounting for Eqs.~\eqref{pi12} and \eqref{ritus_functions}, we obtain
\begin{equation}
\label{dMIII_meaning}
\delta\mathcal{M}^{(\mathrm{III})}-\mathcal{M}^{(1)}\approx i\frac{13m^2g^2\left(\ln{g}-\tilde{C}^{\rm (III)}\right)}{12\pi\sqrt{3}\chi^{1/3}},
\end{equation}
where we introduced the constant $\tilde{C}^{\rm (III)}\approx 4.65 +0.530i$. This difference demonstrates the effect of resumming the perturbative higher order corrections with $r\ge2$ for $g\gg1$. We reflected this symbolically in the top right cell of Table~\ref{tab:scales}.

\section{Summary and discussion}
\label{sec:summ}
After a detailed analysis of radiative corrections in a CCF of up to 3rd-loop order \cite{narozhny1969propagation, ritus1970radiative, ritus1972radiative, morozov1975elastic, morozov1977elastic, narozhny1979radiation, narozhny1980expansion}, Ritus and Narozhny conjectured that in the strong-field regime $\chi\gg 1$ the expansion parameter of QED perturbation theory in a CCF is $g=\alpha\chi^{2/3}$. Recent suggestions \cite{yakimenko2019prospect, baumann2019probing, di2019testing, baumann2019laser, blackburn2019reaching} how this regime could be reached experimentally renewed the interest in this old but so far unsolved problem of quantum field theory. 

The parameter $g$ appears already in the leading-order 1-loop calculation of the correction to the electron mass $\mathcal{M}^{(0)}=\mathcal{O}(g)$ \cite{ritus1972radiative}, and its importance was substantiated further in Ref.~\cite{narozhny1980expansion} by comparing the leading contributions in 2nd and 3rd loop order [see diagrams (2b) and (3g) in Table~\ref{tab:known}]. This analysis suggested that $g$ might be the relevant expansion parameter, i.e., that an all-order non-perturbative resummation becomes necessary in the regime $g\gtrsim 1$. In order to elucidate the Ritus-Narozhny conjecture, we have considered here the high-$\chi$ asymptotic behavior of a certain class of radiative corrections to the electron mass beyond 3 loops, namely the bubble-type corrections to the mass operator shown in Fig.~\ref{fig:diagram}.

The calculation of polarization corrections in a CCF naturally introduces an effective charge $\alpha_{\text{eff}}(l^2,\chi_l)=Z(l^2,\chi_l)\alpha$. It depends both on the photon virtuality $l^2$ and the effective field strength $\chi_l$ (which also scales with the energy of the participating photon). Its dependence on $\chi_l$ turns out to be logarithmic, as one might expect based on the logarithmic effective charge obtained in field-free QED \cite{artimovich1990properties}. However, a strong difference with respect to field-free QED is observed, for example, in the scaling of the mass correction $\mathcal{M}(\chi)$ itself. 

\begin{figure*}
	\centering
	\includegraphics[width=\textwidth]{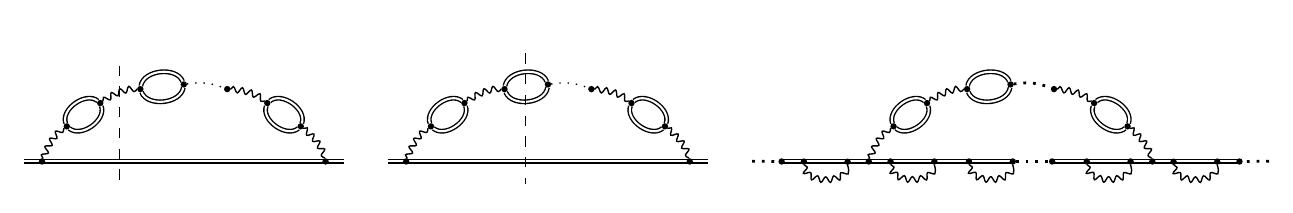}
	\caption{The cuts of the bubble diagram for corrections to photon emission (left) and to trident pair production (center). Right: additional dressing due to electron mass corrections.}
	\label{fig:2cuts}
\end{figure*}

Our findings are summarized in Table~\ref{tab:scales}. The formation scales of the leading $(r+1)$-loop mass correction $\mathcal{M}^{(r)}$ with $r\ge1$ bubble insertions differ from the scales for $r=0$ (no bubbles) and perturbatively are defined by the condition that the polarization operator eigenvalues are not suppressed. In particular, for $\chi\gg 1$, the photon virtuality at higher loop orders, which contains polarization insertions, is much smaller and the associated spatio-temporal extension is much larger than for a loop without such insertions at all. 

According to our analysis, the leading $(r+1)$-loop mass correction scales as $\mathcal{M}^{(r)}=\mathcal{O}(\chi^{-1/3} g^{r+1})$ in a CCF. This is precisely what is asserted in the Ritus-Narozhny conjecture, hence proves it for higher orders $r\ge3$ not considered previously. Notably, the parameter $g=\alpha\chi^{2/3}$ originates for $r\ge2$ from the ratio of the field-induced polarization operator eigenvalue to the photon virtuality $l^2$, evaluated at their typical scales [see Eq.~\eqref{g_meaning}]. The two lowest-order cases $r=0$ (no bubbles) and $r=1$ (single bubble) are special, in particular with respect to their scales. In effect, however, as compared to the above scaling, $\mathcal{M}^{(0)}=\mathcal{O}(g)$ doesn't acquire the factor $\chi^{-1/3}$, whereas $\mathcal{M}^{(1)}$ acquires just an extra factor $\log{\chi}$. The additional factor $\chi^{-1/3}$, arising at higher orders $r\ge1$ due to a modification of the loop formation scale, explains the puzzling anomalous ratio of the 2nd to the 1st loop result [see (1b), (2b) in Table~\ref{tab:known} and Section~\ref{subsec:pert}]. It is worth stressing that all higher-order bubble-type contributions become of the same order for $g\sim 1$. This unambiguously manifests a breakdown of perturbation theory, even if the higher-order contributions remain smaller than the leading-order 1-loop prediction. Therefore, one has to carry out an all-order resummation of such bubble-type contributions for  $g\gtrsim1$.

Here, we study the mass correction $\mathcal{M}(\chi)$ in the regime $g\gtrsim 1$ (see Section~\ref{sec:massop} for the exact definition). The following decomposition is convenient
\begin{gather}
\label{M_decomp}
\mathcal{M}(\chi)=\mathcal{M}^{(0)}(\chi) + \delta{\cal M}, \\
\label{M_decomp1}
\delta{\cal M}=\delta{\cal M}^{(\mathrm{I})}(\chi)+\delta{\cal M}^{(\mathrm{II})}(\chi)+\delta{\cal M}^{(\mathrm{III})}(\chi),
\end{gather}
where $\mathcal{M}^{(0)}(\chi)$ [see Eq.~\eqref{M0}] is the leading-order perturbative result and $\delta{\cal M}$ [see Eqs. \eqref{M_split}, \eqref{deltaM}, and \eqref{deltaM_double}] has been determined by resumming all polarization corrections with $r\ge1$ bubbles, see Fig.~\ref{fig:diagram} and Eq.~\eqref{photon_prop}. Its splitting [as in Eq.~\eqref{M_decomp1}] is stipulated by the composition of the integrand in Eq.~\eqref{deltaM}, namely the terms $\delta{\cal M}^{(\mathrm{I})}$ and $\delta{\cal M}^{(\mathrm{III})}$ correspond to the first term in the integrand, whereas $\delta{\cal M}^{(\mathrm{II})}$ corresponds to the second one. It is convenient to evaluate them separately. 

Notably, the integrand of our non-perturbative result given in Eq.~\eqref{deltaM_double} includes the polarization operator eigenvalues in the exponentials. It turns out that $\delta{\cal M}^{(\mathrm{I})}$ can be neglected (see Section~\ref{sec:dMI}) and that the dominant contributions originate from $\delta{\cal M}^{(\mathrm{II})}$ (see Section~\ref{sec:dMII}) and $\delta{\cal M}^{(\mathrm{III})}$ (see Section~\ref{sec:dMIII}). For them, unlike for $\delta{\cal M}^{(\mathrm{I})}$, the effective formation scales arising during integration, are modified with respect to the perturbative case by involving the coupling $\alpha$. This manifests another aspect of the non-perturbativity of our results. In particular, the effective value of the photon virtuality here corresponds to the bubble-chain dressed photon mass shell, see the last row of Table~\ref{tab:scales}. More generally, the scales of $\delta{\cal M}^{(\mathrm{II})}$ depend on $\chi$ in the same way as the scales of $\mathcal{M}^{(0)}$. However, the spatio-temporal scales are amplified by inverse powers of $\alpha$ and the energy-momentum scales are reduced accordingly. This is consistent with the modification of the perturbative scales at higher orders $r\ge2$, which has been mentioned above. The same correspondence is observed by comparing the scales of $\delta{\cal M}^{(\mathrm{III})}$ with the scales of $\mathcal{M}^{(1)}$.

The importance of our analysis of the formation scales is confirmed by the fact that the asymptotic scalings of $\delta\mathcal{M}^{(\mathrm{II})}$ and $\delta\mathcal{M}^{(\mathrm{III})}$  at $g\gg 1$ [see Eqs.~\eqref{deltaM_II_approx} and \eqref{deltaM_III_approx_final}] can be both understood in a uniform way, as $\delta\mathcal{M}\sim \alpha/s$ with the appropriate choices of $s$ (see Table~\ref{tab:scales}). They can be alternatively represented in terms of other pairs of the three parameters $g$, $\alpha$ and $\chi$, related by our definition $g=\alpha\chi^{2/3}$,
\begin{equation}
\label{M_scaling_final}
\begin{aligned}
\delta\mathcal{M}^{(\mathrm{II})}=\mathcal{O}(\sqrt{\alpha}\, g) =\mathcal{O}(\chi^{-1/3} g^{3/2}) , \\ \delta\mathcal{M}^{(\mathrm{III})} = \mathcal{O}(\sqrt{\alpha}\, g^{3/2}) = \mathcal{O}(\chi^{-1/3} g^{2}).
\end{aligned}
\end{equation}
This result confirms that the parameter $g=\alpha\chi^{2/3}$ determines the scaling of radiative corrections even in the regime $g\gtrsim 1$, where perturbation theory is no longer valid. It is worth noting that a non-analytic dependence on the coupling [e.g., a half-integer power in case of $\delta\mathcal{M}^{(\mathrm{II})}$ or involving a logarithm in case of $\delta\mathcal{M}^{(\mathrm{III})}-\mathcal{M}^{(1)}$, as implied in Eq.~\eqref{dMIII_meaning}], shows that our result is non-perturbative, as it cannot be represented by a power series in the coupling.

As the formation regions differ for $\delta\mathcal{M}^{(\mathrm{II})}$ [see Eq.~\eqref{deltaM_II_approx}] and $\delta\mathcal{M}^{(\mathrm{III})}$ [see Eq.~\eqref{deltaM_III_approx_final}], their physical interpretation should differ as well. According to the optical theorem radiative corrections are closely related to the total probabilities of associated branching processes \cite{berestetskii1982quantum}. The imaginary part of the mass operator determines the electron lifetime inside a background field~\cite{ritus1970radiative,morozov1975elastic,morozov1977elastic,ritus1972radiative,meuren2011quantum}. The electron state can either decay by emitting a photon or by directly producing an electron-positron pair (trident process). Both processes are qualitatively different, in particular with respect to their associated
observables, and are obtained by two types of cuts shown in Fig.~\ref{fig:2cuts}. 

Based on their scaling with $\chi$ we have to associate $\delta\mathcal{M}^{(\mathrm{II})}$ with photon emission and $\delta\mathcal{M}^{(\mathrm{III})}$ with trident pair production, which exhibit the same scaling as the contributions (1b) and (2b) in Table~\ref{tab:known}. This identification is supported by the abnormal and normal signs of the imaginary parts of the corrections $\delta\mathcal{M}^{(\mathrm{II})}$ and $\delta\mathcal{M}^{(\mathrm{III})}$, respectively. The fact that the probability of being in a one-particle state must decay and cannot increase with time determines the allowed total sign of the imaginary part. Therefore, $\delta\mathcal{M}^{(\mathrm{II})}$ must be a correction to the leading-order result $\mathcal{M}^{(0)}$, which clearly describes photon emission. The contribution $\delta\mathcal{M}^{(\mathrm{III})}$, however, has the right sign and describes a decay process which requires at least two interactions, i.e., trident pair production. To leading order the latter process is described\footnote{More precisely, according to \cite{ritus1972vacuum} $\mathrm{Im}\,\mathcal{M}^{(1)}$ contains two contributions. The first one corresponds to the trident process at tree level, and the second one to the interference between the tree level photon emission and 1-bubble correctoin to it. However, the former dominates at $\chi\gg 1$.} by $\mathcal{M}^{(1)}$ \cite{ritus1972radiative, ritus1972vacuum, torgrimsson2020nonlinear} and the non-perturbative correction to it is asymptotically given by Eq.~\eqref{dMIII_meaning}.

\begin{figure}
	\begin{center}
		\includegraphics[width=\columnwidth]{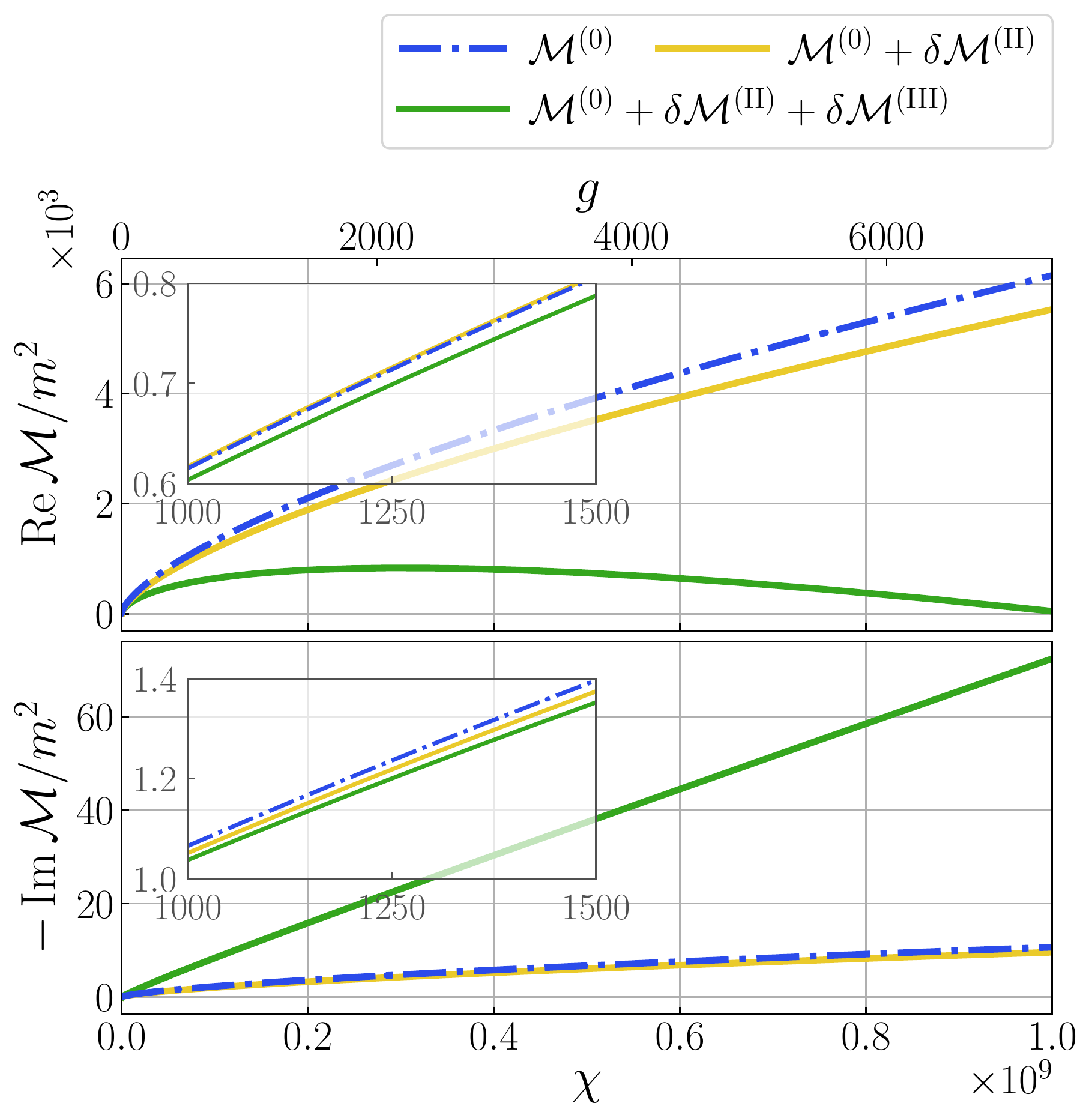}
	\end{center}
	\caption{Dependence of the resummed bubble-type mass correction on $\chi$ (the corresponding value for $g$ is given on the upper axis): the cumulative mass corrections for $\alpha\chi^{2/3}\gg 1$ (inset: the same dependence on $\chi$ magnified in the range near $\alpha\chi^{2/3}\sim 1$). }
	\label{fig:mass_corr}
\end{figure}

The real and imaginary parts of the on-shell mass operator are shown in Fig.~\ref{fig:mass_corr}. The solid yellow line $\mathcal{M}^{(0)}(\chi)+\delta\mathcal{M}^{(\mathrm{II})}(\chi)$ and the dash-dot blue line $\mathcal{M}^{(0)}(\chi)$ demonstrate the impact of polarization effects on photon emission. In the asymptotic region ($g\gg 1$) nonperturbative effects are responsible for a $\sqrt{\alpha}\simeq 10\%$ reduction of both the real and the imaginary part of the invariant amplitude.

In general, however, the contribution $\delta\mathcal{M}^{(\mathrm{III})}(\chi)$ (solid green curve) totally dominates and results in a rather substantial suppression of the real part and an enhancement of the magnitude of the imaginary part. The region $g = \alpha\chi^{2/3}\simeq 1$, which could be accessed experimentally in the mid-term future \cite{yakimenko2019prospect, baumann2019probing, di2019testing, baumann2019laser, blackburn2019reaching}, is shown separately in the insets. The curves have been obtained by a direct numerical evaluation of the integrals in Eqs.~\eqref{deltaM_II} and \eqref{deltaM_III}. In this regime higher-order corrections to photon emission are at the level of $\sim{}0.1\%$ for the real and $\sim{}1\%$ for the imaginary part, respectively. 

We emphasize that the relative smallness of $\delta\mathcal{M}^{(\mathrm{II})}$ and/or $\delta\mathcal{M}^{(\mathrm{III})}$ with respect to $\mathcal{M}^{(0)}$ for $g\simeq 1$ does not imply that the breakdown of perturbation theory is somehow shifted to higher values of $g$. As discussed above, it occurs when all higher-order corrections become of the same order, which happens for $g\sim 1$. The observed suppression is specific to processes like elastic scattering or photon emission. On the other hand, corrections to the trident process included into $\delta\mathcal{M}^{(\mathrm{III})}$, are obviously of the same order as the process itself at the point of breakdown $g\gtrsim 1$ [see Eq.~\eqref{dMIII_meaning} or cf. Eqs.~\eqref{ritus_eq76_bdp} and \eqref{deltaM_III_approx_final}]. We expect the same to be true for general higher-order QED processes. Therefore, our calculations could be tested experimentally, as the regime $g\gtrsim 1$ is accessible in the mid-term future \cite{yakimenko2019prospect, baumann2019probing, di2019testing, baumann2019laser, blackburn2019reaching}. 

Finally, we would like to point out that we only considered one particular subset of diagrams. Hence, further studies are necessary before final conclusions can be drawn. In particular, it should be shown directly that the bubble-type corrections considered here represent indeed the dominant contribution in the asymptotic regime. This dominance is related to an expected suppression of the vertex correction. Whereas this suppression has been proven rigorously in the case of a supercritical magnetic field~\cite{gusynin1999dynamical}, the late work of the Ritus group on this subject actually questioned this assumption for a CCF \cite{morozov1981vertex}. Therefore, the calculation presented in \cite{morozov1981vertex} should be revisited. Naturally, also the electron mass corrections should be resummed, see right panel in Fig.~\ref{fig:2cuts}. Their relative suppression at 3-loop [see diagram (3e) in Table~\ref{tab:known}] could be peculiar to this order. The observed dominance of $\delta\mathcal{M}^{(\mathrm{III})}$ over $\delta\mathcal{M}^{(\mathrm{II})}$ may indicate that other corrections (e.g., rainbow diagrams) with higher multiplicity in the virtual channel are equally or even more important. Furthermore, the direct evaluation of polarization corrections to photon emission and trident pair production would be instructive. Whereas the calculation presented here reveals how the total probabilities scale, modifications to the spectra of branching processes are most easily accessible experimentally.

\textit{Note added}.--- While revising this manuscript, we have noticed a new paper \cite{di2020one}, which generalizes the results of Ref.~\cite{morozov1981vertex} to the plane-wave case. In particular, it confirms the scaling $\mathcal{O}(g)$ of the one-loop vertex correction in a CCF with an on-shell electron and commensurable values of the electron and photon $\chi$-parameters. This, however, is still insufficient for proving or disproving the dominance of the bubble chains, because, as we have seen above, in higher orders the $\chi$-parameters can be effectively distributed non-uniformly. Further investigations of the diagrams containing  vertex corrections are required to ultimately clarify this aspect of the Ritus-Narozhny conjecture. 

\begin{figure*}
	\centering\includegraphics[width=1\textwidth]{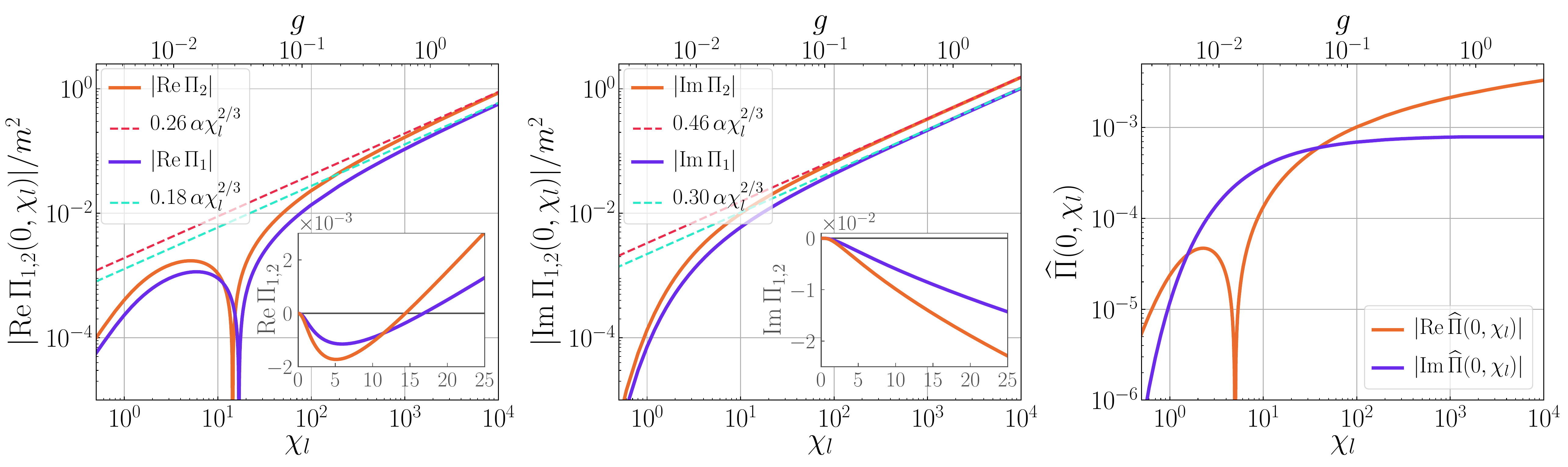}
	\caption{\label{fig:polop_chi} Dependence on $\chi_l$ (the corresponding value for $g$ is given on the upper axis) of the real (left) and imaginary (center) parts of the polarization operator eigenvalues $\Pi_{1,2}$, evaluated on the bare mass shell $l^2=0$, along with the corresponding asymptotics \eqref{polop_asympt} (insets: the same dependence on $\chi_l$ magnified in the range near $\chi_l\sim 1$). Right: the same dependence for the magnitude of the real and imaginary parts of $\widehat{\Pi}=1-Z^{-1}$.}
\end{figure*}

\acknowledgments 
We are grateful to the participants of the specially dedicated meeting ``Physics Opportunities at a Lepton Collider in
the Fully Nonperturbative QED Regime'' (SLAC, 7-9 August, 2019) for valuable discussions.
AAM and AMF were supported by the MEPhI Academic Excellence Project (Contract No. 02.a03.21.0005), Foundation for the advancement of theoretical physics and mathematics ``BASIS'' (Grant No. 17-12-276-1), Russian Foundation for Basic Research (Grants Nos. 19-02-00643, 19-32-60084 and 20-52-12046), and the Tomsk State University Competitiveness Improvement Program. At Princeton, SM received funding from the Deutsche Forschungsgemeinschaft (DFG, German Research Foundation) under Grant No. 361969338. At Stanford, SM was supported by the U.S. Department of Energy under contract number DE-AC02-76SF00515.

%\clearpage

\appendix*

%\clearpage
\section{One-loop polarization operator in a constant crossed field}
\label{sec:polop}

\begin{figure}
	\centering\includegraphics[width=\columnwidth]{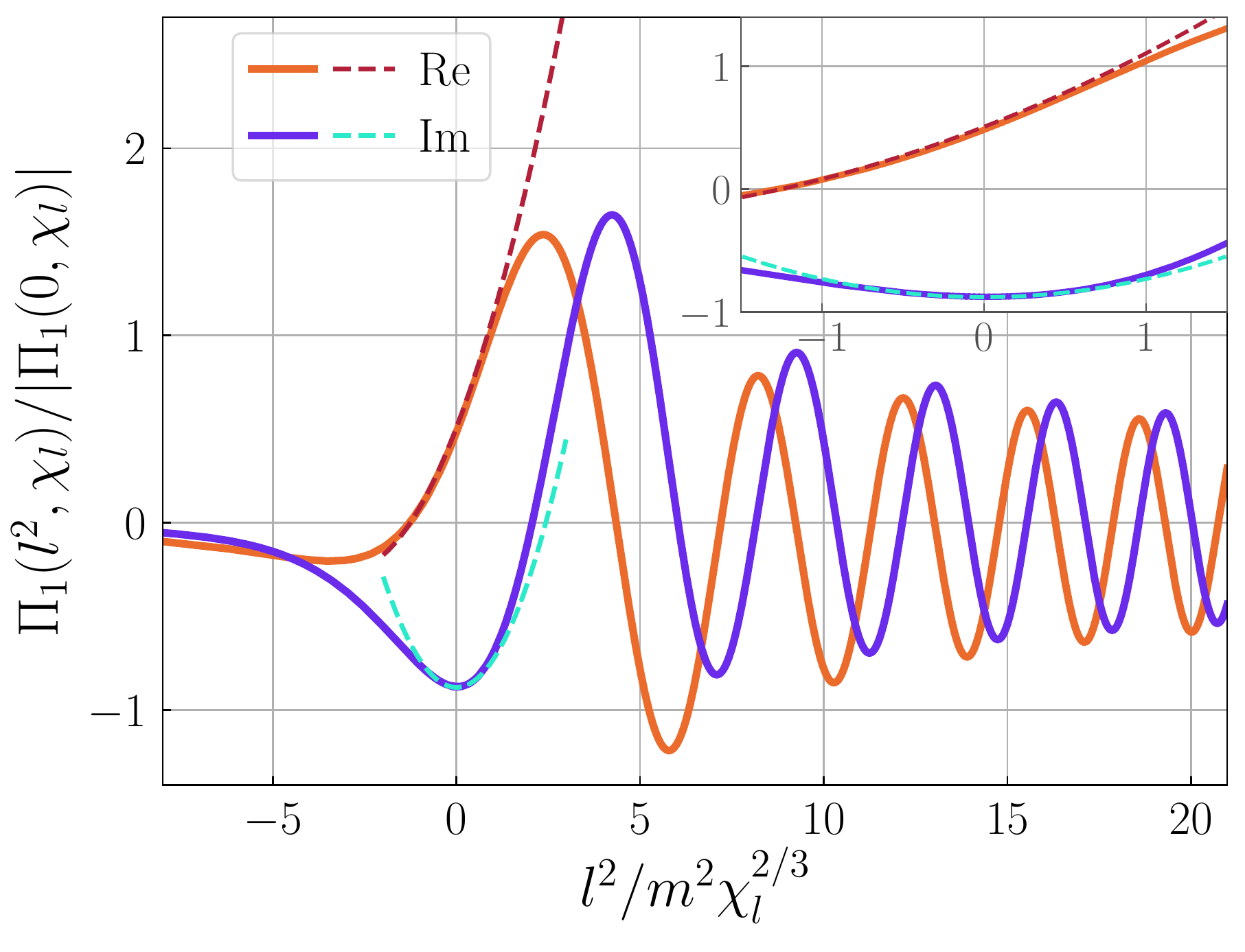}
	\caption{\label{fig:polop_l2} Dependence on $l^2$ of the real and imaginary parts of the polarization operator eigenvalue $\Pi_1(l^2,\chi_l=10^4)$: asymptotic expression given in Eq.~\eqref{polop_asympt_off_shell} (dashed lines) vs direct numerical evaluation of Eq.~\eqref{pi12} (solid lines). The axes are scaled such that the curves are stable under changing of $\chi_l$.}
\end{figure}

For completeness, we provide the explicit expressions for the renormalized one-loop polarization operator in a CCF \cite{narozhny1969propagation,ritus1970radiative,ritus1985quantum}
\begin{equation}\label{polop_struct}
\begin{split}
\Pi_{\mu\nu}(l) &= \widehat{\Pi}(l^2,\chi_l) \left(l^2 g_{\mu\nu}-l_\mu l_\nu\right)\\&+\sum\limits_{i=1}^2\Pi_i(l^2,\chi_l)\epsilon_\mu^{(i)}(l)\epsilon_\nu^{(i)}(l),
\end{split}
\end{equation}
where the vectors $\epsilon_\mu^{(1)}(l)=eF_{\mu\nu}l^\nu/(m^3\chi_l)$ and $\epsilon_\mu^{(2)}(l)=eF^\star_{\mu\nu}l^\nu/(m^3\chi_l)$ are the same as in Eq.~\eqref{photon_prop}.
Its three non-trivial renormalized eigenvalues read
\begin{equation}\label{pi12}
\Pi_{1,2}(l^2,\chi_l)=\frac{4\alpha\chi_l^{2/3} m^2}{3\pi}\int_4^\infty \frac{dv}{v^{13/6}} \,\frac{v+0.5\mp1.5}{\sqrt{v-4}}f'(\zeta),
\end{equation}
and
\begin{equation}\label{pi3}
\begin{split}
l^2\widehat{\Pi}(l^2,\chi_l)=-l^2\frac{4\alpha}{\pi} & \int_4^\infty \frac{dv}{v^{5/2}\sqrt{v-4}}\\
& \times \left[f_1(\zeta)-\log\left(1-\frac{1}{v}\frac{l^2}{m^2}\right)\right].
\end{split}
\end{equation}
Here
\begin{equation}\label{zeta}
\zeta=\left(\frac{v}{\chi_l}\right)^{2/3}\left(1-\frac{l^2}{vm^2}\right),
\end{equation}
is the argument of the Ritus functions
\begin{equation}\label{ritus_functions}
\begin{aligned}
f(\zeta)=i\int_0^\infty d\sigma\, e^{-i(\zeta\sigma+\sigma^3/3)},\\
f_1(\zeta)=\int_{\zeta}^\infty dz \left[f(z)-\frac{1}{z}\right],
\end{aligned}
\end{equation}
which are defined as in \cite{ritus1985quantum}, and $f'(\zeta)$ is the derivative of the former.

%\begin{equation}\label{ritus_functions}
%\begin{aligned}
%f(\zeta)=i\int_0^\infty d\sigma\, e^{-i(\zeta\sigma+\sigma^3/3)},\\
%f_1(\zeta)=\int_{\zeta}^\infty dz \left[f(z)-\frac{1}{z}\right]
%\end{aligned}
%\end{equation}
%are the Ritus functions, which are defined as in \cite{ritus1985quantum}, and $f'(\zeta)$ is the derivative of the former one, and their argument is
%\begin{equation}\label{zeta}
%\zeta=\left(\frac{v}{\chi_l}\right)^{2/3}\left(1-\frac{l^2}{vm^2}\right).
%\end{equation}
When the external field is switched off, $\Pi_{1,2}$ vanish and $l^2\widehat{\Pi}$ is reduced to the well-known expression for the one-loop polarization operator in field-free QED \cite{peskin2018introduction}. We assume the standard renormalization condition that the expressions in Eqs.~\eqref{pi12}, \eqref{pi3} vanish at $l^2=0$ in the absence of the external field (for $\chi_l=0$) \cite{ritus1972radiative}.

By carrying out a resummation of the Dyson series one obtains the following expression for the bubble-chain photon propagator \cite{ritus1972radiative}
\begin{equation}\label{D0_rep}
\begin{split}
D^c_{\mu\nu}(l)= & \frac{-i g_{\mu\nu}}{l^2-l^2\widehat{\Pi}}\\
&+\sum\limits_{i=1}^2\frac{i\Pi_i}{(l^2-l^2\widehat{\Pi})(l^2-l^2\widehat{\Pi}-\Pi_i)}\epsilon_\mu^{(i)}(l)\epsilon_\nu^{(i)}(l).
\end{split}
\end{equation}
With the notation 
\begin{equation}\label{Z}
Z(l^2,\chi_l)=\frac{1}{1-\widehat{\Pi}(l^2,\chi_l)}
\end{equation}
the propagator $D^c_{\mu\nu}(l)$ takes the form given in Eq.~\eqref{photon_prop}. In any diagram the propagator always connects two vertices. Therefore, the factor $Z$ appears only in combination with $\alpha$. Together they compose the effective coupling $\alpha_{\text{eff}}(l^2,\chi_l)=Z(l^2,\chi_l)\alpha$. We adopt the terminology of Ref.~\cite{artimovich1990properties}, where the value $\alpha_{\text{eff}}(0,\chi_l)$ is called the field-dependent effective charge.

Note that for $\chi_l\gg 1$ and for the bare on-shell condition $l^2=0$ we have
\begin{equation}\label{polop_asympt}
\begin{aligned}
\widehat{\Pi}(0,\chi_l)\simeq\frac{\alpha}{3\pi}\log{\chi_l^{2/3}},\\
\Pi_i(0,\chi_l) = \alpha m^2\pi_i(\chi_l),\\
\pi_i(\chi_l)\simeq K_i\chi_l^{2/3},
\end{aligned}
\end{equation}
where
\begin{equation}\label{K_i}
%K_1=0.175(1-i\sqrt{3}),\quad K_2=1.5 K_1.
K_{1,2}= e^{-i\pi/3}\frac{5\mp1}{6^{4/3}\sqrt{\pi}}\frac{\Gamma^2\left(2/3\right)}{\Gamma\left(13/6\right)}.
\end{equation}
The dependence of the on-shell expressions given in Eqs.~\eqref{pi12} and \eqref{pi3} on $\chi_l$ is shown in Fig.~\ref{fig:polop_chi}. One can see that the asymptotics given in Eq.~\eqref{polop_asympt} are achieved for $\chi_l\gtrsim 10^3$ and that $\widehat{\Pi}=\mathcal{O}(10^{-2})$ for all reasonable values of $\chi_l$. Since asymptotically $\widehat{\Pi}(l^2,\chi_l)=\mathcal{O}(\alpha)$ has only a weak logarithmic dependence on $\chi_l$ and $l^2$, it is possible to neglect small modifications of the effective charge by setting $Z(l^2,\chi_l)\approx 1$ and $\alpha_{\text{eff}}\approx \alpha$ throughout the paper.

The off-shell dependence of $\Pi_1(l^2,\chi_l)$ on $l^2/m^2\chi_l^{2/3}$ is shown in Fig.~\ref{fig:polop_l2}. One can see that it decays exponentially to the left of the origin and exhibits a power law decay at the same scale as it oscillates to the right.
%it expenentially decays to the left and fades out oscillating to the right. 
The optimal values are acquired near the bare mass shell (for $|l^2|\lesssim m^2\chi_l^{2/3}$), where one can expand Eq.~\eqref{pi12} into powers of the virtuality $l^2$,
\begin{equation}\label{polop_asympt_off_shell}  
\begin{split}       
\Pi_i(l^2,\chi_l)\approx m^2\alpha\chi_l^{2/3}\left[K_i\vphantom{\left(\frac{l^2}{m^2\chi_l^{2/3}}\right)^2}\right.&+K_i^{(1)}\frac{l^2}{m^2\chi_l^{2/3}}\\
&\left.+K_i^{(2)}\left(\frac{l^2}{m^2\chi_l^{2/3}}\right)^2\right],
\end{split}
\end{equation}
where
\begin{equation}\label{polop_asympt_off_shell_K} 
\begin{aligned}
K_i^{(1)}=&\frac{13\mp3}{18\pi},\\
K_i^{(2)}=&e^{i\pi/3}\frac{4\mp1}{4\cdot6^{2/3}\sqrt{\pi}}\frac{\Gamma^2\left(4/3\right)}{\Gamma\left(17/6\right)}.
\end{aligned}
\end{equation}
Note that the off-shell correction linear in $l^2$ is real. One can see from Fig.~\ref{fig:polop_l2} that the asymptotics given in Eq.~\eqref{polop_asympt} remains a good order-of-magnitude estimate even for $|l^2|\lesssim m^2\chi_l^{2/3}$.

Finally, in virtue of Eqs.~\eqref{pi12}, \eqref{zeta} and \eqref{ritus_functions}, $\Pi_i$ can be represented by a one-sided Fourier integral [see Eq.~\eqref{pi_fourier}], where
\begin{equation}\label{pi_Fourier}
\begin{split}
\tilde{\Pi}_{1,2}(\tau,\chi_l)=\frac{4\alpha}{3\pi}\chi_l^2\tau m^6 & \int_4^\infty  \frac{dv}{v^{3/2}} \,\frac{v+0.5\mp1.5}{\sqrt{v-4}}\\
& \times e^{-im^2v(\tau+m^4\chi_l^2\tau^3/3)},
\end{split}
\end{equation}
and the characteristic values of the variables around which the integral is formed are obviously $v\simeq 1$ and $\tau\simeq \min\left\{m^{-2},m^{-2}\chi_l^{-2/3}\right\}$.

%\clearpage

%\bibliography{lit}


\end{document}